\documentclass{aastex631}

\newcommand{\ulx}{SS\,433}
\newcommand{\ms}{\,M$_\odot$}
\newcommand{\erg}{\,erg}
\newcommand{\kyr}{\,kyr}
\newcommand{\yr}{\,yr}
\newcommand{\km}{\,km}
\newcommand{\cm}{\,cm}
\newcommand{\pc}{\,pc}
\newcommand{\kpc}{\,kpc}
\newcommand{\keV}{\,keV}
\newcommand{\gev}{\,GeV}
\newcommand{\tev}{\,TeV}
\newcommand{\pcc}{\,cm$^{-3}$}
\newcommand{\ps}{\,s$^{-1}$}
\newcommand{\pyr}{\,yr$^{-1}$}
\newcommand{\psqd}{\,deg$^{-2}$}
\newcommand{\ghz}{\,GHz}
\newcommand{\ks}{\,ks}

\newcommand{\apec}{{\it apec}}
\newcommand{\vap}{{\it vapec}}
\newcommand{\vnei}{{\it vnei}}

\newcommand{\vps}{{\it vpshock}}
\newcommand{\pow}{{\it powerlaw}}
\newcommand{\tb}{{\it TBabs}}

\newcommand{\xmm}{\sl XMM-Newton\rm}
\newcommand{\rosat}{\sl ROSAT\rm}
\newcommand{\vla}{\sl VLA\rm}
\newcommand{\hi}{\sl Acrecibo\rm}
\newcommand{\hess}{H.E.S.S.}

\usepackage{multirow}%
\usepackage{booktabs}
\usepackage{subfigure}
\usepackage{amsmath,amssymb,amsfonts}%
\usepackage{amsthm}%
\usepackage{CJK}

\begin{document}
\begin{CJK*}{UTF8}{gbsn}
\title{An X-ray Shell Reveals the Supernova Explosion for Galactic Microquasar SS~433}

\author[0000-0002-2421-173X]{Yi-Heng Chi (池奕恒)}
%\author[0000-0002-2421-173X]{Yi-Heng Chi}
\affiliation{School of Astronomy \& Space Science, Nanjing University, 163 Xianlin Avenue, Nanjing 210023, China}

\author[0000-0001-8674-2336]{Jiahui Huang (黄佳辉)}
%\author[0000-0001-8674-2336]{Jiahui Huang}
\affiliation{Department of Engineering Physics, Tsinghua University, Beijing 100084, China}
\affiliation{Center for Computational Sciences, University of Tsukuba, Tennodai 1-1-1, Tsukuba, Ibaraki 305-8577, Japan}

\author[0000-0002-5683-822X]{Ping Zhou (周平)}
%\author[0000-0002-5683-822X]{Ping Zhou}
\affiliation{School of Astronomy \& Space Science, Nanjing University, 163 Xianlin Avenue, Nanjing 210023, China}
\affiliation{Key Laboratory of Modern Astronomy and Astrophysics, Nanjing University, Ministry of Education, Nanjing 210023, China}

\author[0000-0001-7584-6236]{Hua Feng (冯骅)}
%\author[0000-0001-7584-6236]{Hua Feng}
\affiliation{Key Laboratory of Particle Astrophysics, Institute of High Energy Physics, Chinese Academy of Sciences, Beijing 100049, China}

\author[0000-0002-0584-8145]{Xiang-Dong Li (李向东)}
%\author[0000-0002-0584-8145]{Xiang-Dong Li}
\affiliation{School of Astronomy \& Space Science, Nanjing University, 163 Xianlin Avenue, Nanjing 210023, China}
\affiliation{Key Laboratory of Modern Astronomy and Astrophysics, Nanjing University, Ministry of Education, Nanjing 210023, China}

\author[0000-0001-9564-0876]{Sera B. Markoff}
\affiliation{Anton Pannekoek Institute for Astronomy, University of Amsterdam, Science Park 904, 1098XH Amsterdam, The Netherlands}
\affiliation{Gravitation Astroparticle Physics Amsterdam (GRAPPA) Institute, University of Amsterdam, \\
Science Park 904, 1098XH Amsterdam, The Netherlands}

\author[0000-0001-6189-7665]{Samar Safi-Harb}
\affiliation{Department of Physics and Astronomy, The University of Manitoba, Winnipeg, MB R3T 2N2, Canada}

\author[0000-0002-9105-0518]{Laura Olivera-Nieto}
\affiliation{Max-Planck-Institut f{\"u}r Kernphysik, P.O. Box 103980, D 69029 Heidelberg, Germany}

\correspondingauthor{Ping Zhou}
\email{pingzhou@nju.edu.cn}

\correspondingauthor{Hua Feng}
\email{hfeng@ihep.ac.cn}
\begin{abstract}

How black holes are formed remains an open and fundamental question in Astrophysics. Despite theoretical predictions, it lacks observations to understand whether the black hole formation experiences a supernova explosion. Here we report the discovery of an X-ray shell north of the Galactic micro-quasar \ulx\ harboring a stellar-mass black hole spatially associated with radio continuum and polarization emissions, and an HI cloud. Its spectrum can be reproduced by a 1-keV under-ionized plasma, from which the shell is inferred to have been created by a supernova explosion 20--30\kyr\ ago and its properties constitute evidence for canonical SN explosions to create some black holes. Our analysis precludes other possible origins including heated by jets or blown by disk winds. According to the lower mass limit of the compact object in \ulx, we roughly deduced that the progenitor should be more massive than 25\ms. The existence of such a young remnant in \ulx\ can also lead to new insights into the supercritical accretion in young microquasars and the $\gamma$-ray emission of this system. The fallback ejecta may provide accretion materials within tens of thousands of years while the shock of the supernova remnant may play a crucial role in the cosmic ray (re)acceleration.

\end{abstract}

\keywords{Supernova remnants (1667) --- Stellar mass black holes (1611) --- X-ray binary stars (1811) --- Ultraluminous x-ray sources (2164) --- Stellar accretion disks (1579)}

\section{Introduction} \label{sec: intro}
The young \citep[$10^4-10^5$\yr,][]{Zealey1980, Shkovskii1981, Heuvel1981, Lockman2007, Goodall2011, Panferov2017} high-mass X-ray binary \ulx\ is a famous system with relativistic jets and may be the only Galactic prototype of an ultra-luminous X-ray source \citep{Brinkman2007, Safi-Harb2022}. 
\ulx\ stands out from many other X-ray binaries due to its supercritical gas accretion onto a compact object, where the mass-accretion rate greatly exceeds the Eddington rate, defined by balancing the radiation pressure with gravity. The mass of the compact object remains uncertain despite decades of study, but the latest results suggest a high mass ratio of the compact object to the donor star of $q\gtrsim0.6$ and a compact object mass of 5--15\ms, favoring a stellar black hole over a neutron star \citep{Gies2002, Bowler2018, Cherepashchuk2019}. This energetic system is known to strongly modify its surroundings. \ulx\ is harbored within a giant radio nebula, W50 referred to as the Manatee nebula, with an angular size 2°$\times$1° \citep[174\pc$\times$87\pc\ at a distance of 5\kpc,][]{Su2018} \citep{Dubner1998}, whose major axis from west to east aligns to the X-ray lobes shaped by the relativistic jets (Figure~\ref{fig: dual}). Perpendicular to the collimated jets, W50 still has a remarkable extension of 1°, requiring other mechanisms to drive the bubble, such as an SNR, equatorial disk winds, or different jet episodes in the past \citep{Fabrika2004, Goodall2011, Farnes2017}. The physical nature of W50 thus links some crucial questions on stellar-mass black holes, the most extreme stellar objects in our Universe: Was the black hole created with an SN explosion? What is the feedback and evolution of a microquasar? Past observations focused primarily on the bright non-thermal X-ray lobes which probe the impact of the jets with the W50 nebula in the east-west direction \citep{Brinkman2007, Safi-Harb2022}. With the new \xmm\ observations, we target the perpendicular direction where other mechanisms may stand out free from contamination by the luminous jets. As such, we conduct the first search and study of the X-ray emission in the northern region of W50 and outside the precession cone of \ulx. 

This letter is organized as follows: we present our analysis based on \xmm\ observations and the newly discovered X-ray shell in \S\ref{sec: data} and \S\ref{sec: result}, respectively. Then we discuss its origin as an SNR (\S\ref{sec: ori}), the mass of the supernova progenitor (\S\ref{sec: mass}), and the new insights the SNR brings us on the supercritical accretion (\S\ref{sec: acc}) and $\gamma$-ray emission of \ulx/W50 system (\S\ref{sec: tev}).

\section{XMM-Newton observations and data reduction}
\label{sec: data}
The North of W50 was observed by \xmm\ \citep{Jansen2001} in Oct.\ 2021 and Apr.\ 2022 (ObsID: 0882560201 and 0882560101, PI: P. Zhou). Observation 0882560201 was heavily contaminated by soft proton flares, leading to a reduced flare-free exposure time of 19.6, 23.5, and 9.6\ks\ for MOS1 \citep{Tuner2001}, MOS2, and pn camera \citep{Struder2001}, respectively. We mainly used observation 0882560101 pointing to W50 Northeast, whose flare-filtered exposure is 40.9, 40.4, and 33.2\ks\ for three instruments. We retrieved archival \xmm\ data of \ulx\ (0694870201, PI: A. Medvedev) and the luminous jet lobes (0840490101, PI: S. Safi-Harb; 0075140401 and 0075140501, PI: W. Brinkmann) for a large-scale X-ray image of W50. We reproduced and analyzed \xmm\ observations with the \xmm\ Science Analysis System \citep[SAS, version 20.0,][]{sas} and the \xmm\ Extended Source Analysis Software (XMM-ESAS, version 0.11.2) for the background modeling \citep{Snowden2004}. All the raw data are reprocessed by {\it emchain} and {\it epchain}, and then the contamination of flares is filtered using {\it mos-filter} and {\it pn-filter}. 

\section{Results}
\label{sec: result}
\subsection{Morphology}
\begin{figure*}
    \centering
    \includegraphics[width=\textwidth]{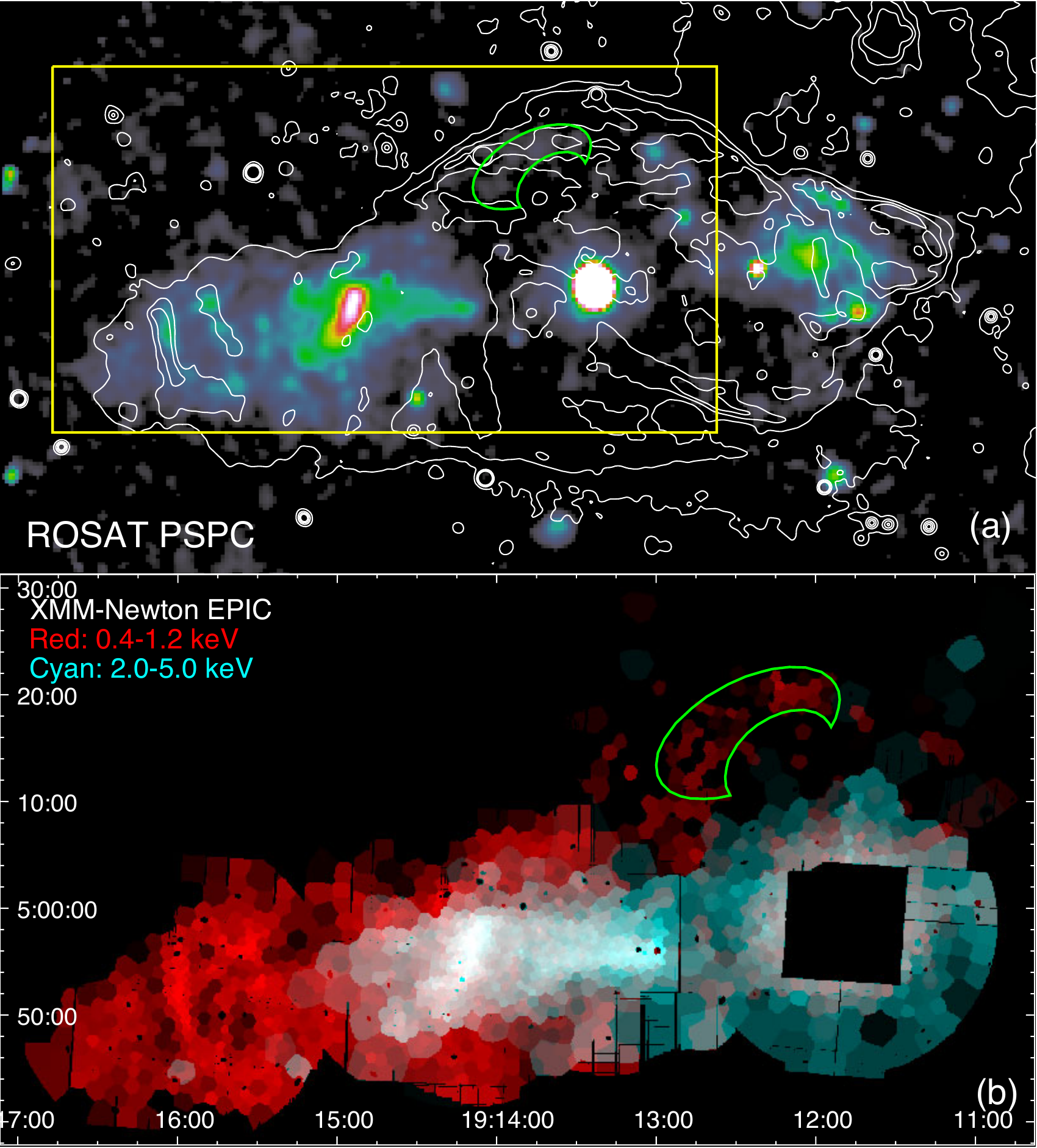}
    \caption{\small Location of the X-ray shell. {\bf (a)}: \rosat\ PSPC X-ray image (see Appendix~\ref{app: rosat}) of the nebula W50 overlapped by the intensity contour of the 1.4\ghz\ \vla\ observation \citep{Dubner1998}. {\bf (b)}: Zoom-in dual-band \xmm\ image in the region marked in {\it yellow} in 0.4--1.2\keV\ ({\it red}) and 2.0--5.0\keV\ ({\it cyan}) with point-like sources subtracted. Each bin shares an average signal-to-noise ratio of 10. The softer and harder energy bands highlight the thermal and non-thermal emission of the nebula, respectively. The {\it green} region labels the shell-like structure.}
    \label{fig: dual}
\end{figure*}

We selected the 0.4--1.2\keV\ and 2.0--5.0\keV\ bands for imaging analysis, where the softer band traces thermal emission from hot plasma and where the non-thermal synchrotron emission dominates the harder X-ray band. The intermediate energy band was excluded to avoid the strong instrumental Al~K (1.49\keV, for MOS and pn) and Si~K (1.75\keV, for MOS) lines. For each observation, {\it mos-spectra} and {\it pn-spectra} produce images and exposure maps in the two selected bands for each camera. The instrumental background of each image, or rather, quiescent particle background (QPB), is modeled through {\it mos\_back} or {\it pn\_back} and projected to the sky by {\it rot-im-det-sky}. All the images, exposure maps, and QPB backgrounds were mosaicked using {\it merge\_comp\_xmm}. To underline the faint structure of the diffuse emission, we applied the task {\it cheese} to detect and mask point-like sources in each band and subsequently adaptively binned the image based on Weighted Voronoi Tessellation algorithm \citep{Cappellari2003, Diehl2006} to ensure an average signal-to-noise ratio of 10 per bin. The dual-band false-color image (Figure~\ref{fig: dual}) shows the heated gas (red) and synchrotron emission of the jet (cyan). 

\begin{figure*}
    \centering
    \includegraphics[width=\textwidth]{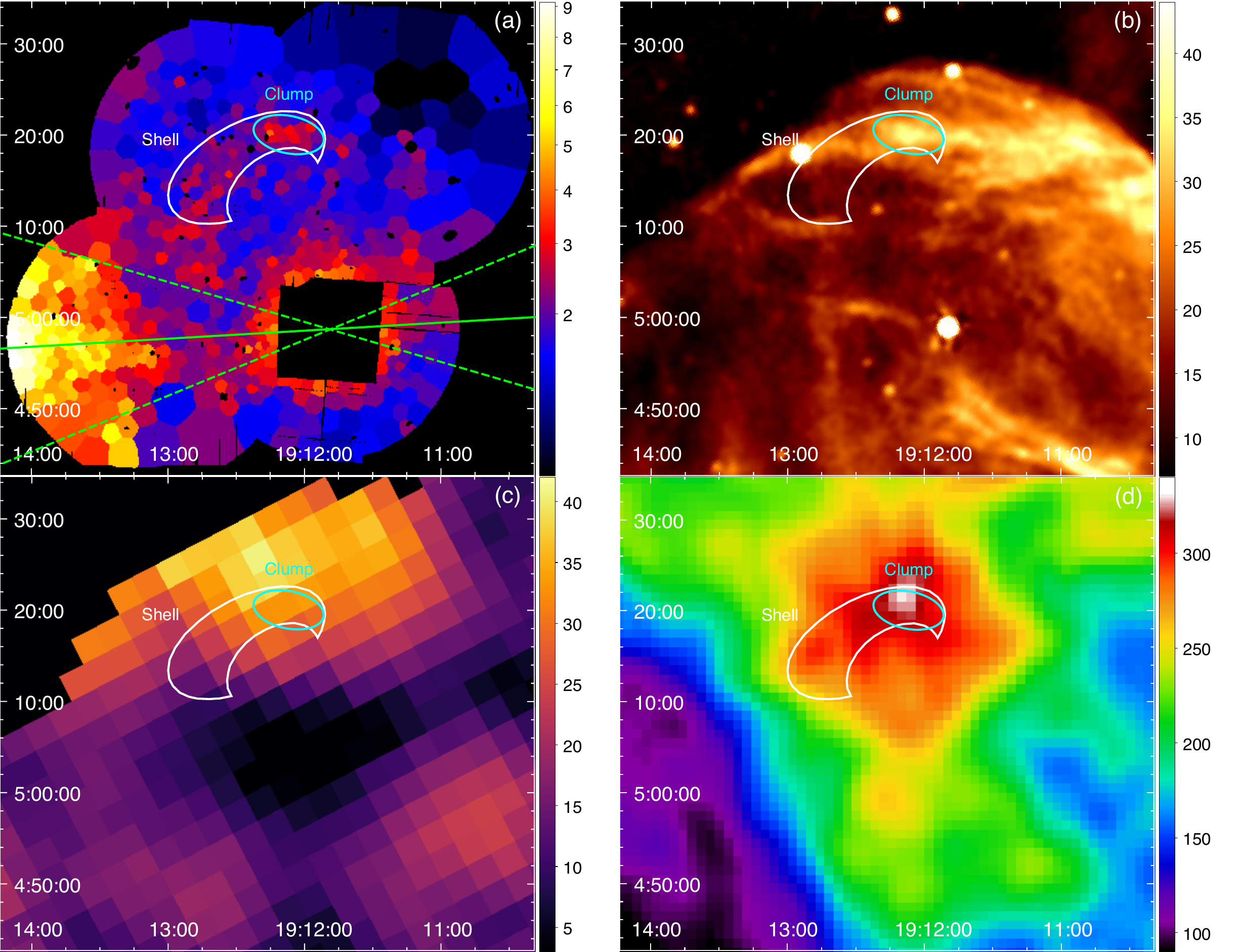}
    \caption{\small Zoom-in multi-band images of our region of interest. {\bf (a)}: \xmm\ EPIC 0.4--1.2\keV\ mosaic in units of counts\psqd\ps. Each bin has a signal-to-noise ratio of $\sim10$. The X-ray shell and clump are labeled in {\it white} and {\it cyan}, respectively. 
    The green solid and dashed lines mark the jet axis and cone of the jet precession and nod, respectively \citep{Fabrika2004}.
    {\bf (b)}: \vla\ 1.4\ghz\ intensity map in units of mJy/beam \citep{Dubner1998}. {\bf (c)}: Sino-German 6\cm\ polarized fraction map in units of percentage. \citep{Gao2011} {\bf (d)}: \hi\ HI intensity in 68--75\km\ps\ in units of K \citep{Peek2018, Su2018}. }
    \label{fig: 4band}
\end{figure*}

A shell-like structure appears northeast of \ulx\ in the 0.4--1.2 keV band (Figure~\ref{fig: dual} \&~\ref{fig: 4band}{\bf (a)}). 
Although fainter than in the jet-impact areas in the east and west, the X-ray structure is clearly detected and spatially correlated with the radio shell in the north. The emission of this structure above 2\keV\ is too weak to be detected in that band, indicative of a soft X-ray spectrum. Figure~\ref{fig: 4band} provides a close-up image of the northern X-ray shell {\bf (a)}, in comparison with the radio emission {\bf (b)}, polarized radio signal {\bf (c)}, and the HI gas {\bf (d)}. The shell has a curvature radius of $\sim20'$ and is roughly centered at the position of \ulx. An X-ray-enhanced clump appears in the north, where the radio emission is locally enhanced \citep{Dubner1998} and highly polarized \citep{Gao2011, Farnes2017}. Meanwhile, the X-ray clump is spatially associated with an HI cloud at a distance of $\sim5$\kpc\ \citep{Su2018}, probably indicative of a density enhancement due to an interaction with the HI cloud. 

\subsection{Spectral Analysis}
We analyze the spectra of the whole northern X-ray shell ($\sim9.4'\times4.9'$, defined as ``Shell'') and the sub-structure clump ($\sim3.9'\times2.0'$, defined as ``Clump''). The source regions are shown in Figure~\ref{fig: 4band} and the sky background is selected outside the radio boundary of W50. 
The instrumental background matters, especially when analyzing weak diffuse emission and when the instrumental background is non-uniform across the field of view.
Therefore, instead of directly subtracting the sky background, we obtained the sky background model by fitting the QPB-subtracted background spectra and applied the model to the source spectra by multiplying an area ratio between the source and background regions. The task {\it mos-spectra} and {\it pn-spectra} extracted spectra and generated response files, while the QPB spectra of each region are modeled by {\it mos\_back} and {\it pn\_back}. The spectra were grouped with {\it ftgrouppha} to guarantee at least 50 counts per bin in favor of chi-square statistics. XSPEC \citep[version 12.13.0c,][]{xspec} with AtomDB (version 3.0.9) was used for spectral analysis. The fitting parameters are listed in Table~\ref{tab: spec} and spectra with folded models are shown in Appendix~\ref{app: spec}.

\begin{table*}
\centering
    \caption{\small XMM-Newton X-ray spectral fitting results. The errors correspond to 1$\sigma$ uncertainties.}
    \renewcommand\arraystretch{1.5}
    \begin{tabular}{ccccccccccc}
        \hline\hline
        \multirow{3}{*}{Region} & Absorption & \multicolumn{8}{c}{plasma} & \multirow{3}{*}{$\chi^2$/d.o.f.} \\ \cmidrule(lr){2-2}\cmidrule(lr){3-10}
        & $N_{\rm H}$ & \multirow{2}{*}{model} & kT & $\tau$ & norm & O & Ne & Mg & Si & \\
        & $10^{21}\rm\,cm^{-2}$ &  & keV & $10^{10}$\,s\,cm$^{-3}$ & $10^{-4}\rm\,cm^{-5}$ & Z/Z$_{\odot}$ & Z/Z$_{\odot}$ & Z/Z$_{\odot}$ & Z/Z$_{\odot}$ & \\ \hline
        \multirow{5}{*}{Shell} & $6.61_{-0.81}^{+0.90}$ & \vnei & $1.23_{-0.33}^{+0.58}$ & $2.6_{-0.9}^{+1.9}$ & $1.27_{-0.31}^{+0.51}$ & $\cdots$ & $\cdots$ & $1.22_{-0.28}^{+0.33}$ & $2.64_{-0.60}^{+0.77}$ & 546.8/467 \\
        & $6.9_{-2.7}^{+1.9}$ & \vps & $1.1_{-0.3}^{+1.0}$ & $5.2_{-3.2}^{+10.0}$ & $1.63_{-0.81}^{+0.98}$ & $0.84_{-0.58}^{+1.02}$ & $\cdots$ & $1.21_{-0.34}^{+0.49}$ & $3.03_{-0.84}^{+0.65}$ & 547.7/466 \\
        & $4.87_{-0.89}^{+0.74}$ & \vap & $0.77_{-0.03}^{+0.04}$ & $\cdots$ & $1.62_{-0.28}^{+0.31}$ & $9.4_{-2.7}^{+3.1}$ & $\cdots$ & $3.3_{-1.1}^{+1.4}$ & $3.5_{-1.0}^{+1.1}$ & 547.4/467 \\ 
        & $5.1_{-1.9}^{+1.2}$ & \vap* & $0.74_{-0.04}^{+0.05}$ & $\cdots$ & $2.07_{-0.74}^{+0.70}$& $\cdots$ & $\cdots$ & $3.4_{-1.1}^{+2.1}$ & $3.5_{-1.0}^{+1.8}$ & 561.7/468 \\
        & $\cdots$ & Null & $\cdots$ & $\cdots$ & $\cdots$ & $\cdots$ & $\cdots$ & $\cdots$ & $\cdots$ & 763.9/472 \\ \hline
        \multirow{5}{*}{Clump} & $10.3_{-1.6}^{+1.9}$ & \vnei & $0.8_{-0.3}^{+1.1}$ & $1.0_{-0.3}^{+4.8}$ & $1.4_{-1.1}^{+2.9}$ & $\cdots$ & $\cdots$ & $1.07_{-0.33}^{+0.39}$ & $2.2_{-1.3}^{+2.4}$ & 195.5/162 \\
        & $8.3_{-2.8}^{+8.1}$ & \vps & $0.92_{-0.70}^{+0.78}$ & $3.1_{-1.7}^{+9.7}$ & $0.8_{-0.5}^{+2.5}$ & $0.6_{-0.4}^{+1.0}$ & $\cdots$ & $1.25_{-0.67}^{+0.64}$ & $1.8_{-1.0}^{+2.7}$ & 194.9/161 \\
        & $4.9_{-1.7}^{+1.6}$ & \vap & $0.74_{-0.07}^{+0.07}$ & $\cdots$ & $0.55_{-0.19}^{+0.30}$ & $5.9_{-3.6}^{+4.6}$ & $<4.5$ & $3.8_{-1.7}^{+2.8}$ & $1.8_{-1.4}^{+2.0}$ & 200.4/161 \\
        & $6.3_{-2.1}^{+1.6}$ & \vap* & $0.71_{-0.06}^{+0.07}$ & $\cdots$ & $0.86_{-0.33}^{+0.41}$ & $\cdots$ & $\cdots$ & $2.9_{-1.2}^{+2.2}$ & $1.5_{-0.9}^{+1.3}$ & 202.5/163 \\
        & $\cdots$ & Null & $\cdots$ & $\cdots$ & $\cdots$ & $\cdots$ & $\cdots$ & $\cdots$ & $\cdots$ & 294.9/167 \\ \hline\hline   
    \end{tabular}
    \label{tab: spec}
\end{table*}

When fitting the sky background, we considered the X-ray emission from the local hot bubble (\apec), Galactic ISM (\vap), and background AGN \citep[\pow\ with a photon index of 1.4,][]{Chen1997} \citep{Snowden2004, Kuntz2008}. To further mimic the spectra, we let the abundances of \vap\ vary and got a reduced chi-square ($\chi^2_r$) of 0.98 with 191 degrees of freedom. For source spectra, apart from the sky background, we tested four different models, including \pow\ (for non-thermal emission), \vap\ \citep[plasma in collision-ionization-equilibrium, CIE,][]{apec}, \vnei\ (non-equilibrium ionization, NEI, plasma), and \vps\ \citep[plane-parallel shocked non-equilibrium plasma,][]{Borkowski2001}. The absorption component of all these models is \tb\ \citep{Wilms2000}.

The non-thermal model gives a high or steep photon index $\Gamma$ of $5.71_{-0.55}^{+0.66}$ for the Shell at a confidence level of 1$\sigma$, much higher than the synchrotron emission detected in the lobes \citep[$\Gamma\approx1.6$,][]{Safi-Harb2022, Brinkman2007} and a typical value of $<3$ for non-thermal objects. In contrast, such a photon index suggests a soft spectrum probably dominated by thermal X-ray emission. Meanwhile, the reduced chi-square is 1.31 with 470 degrees of freedom, significantly poorer than any other model. The NEI models, \vnei\ and \vps, are widely used to describe X-ray plasmas in SNRs \citep{Borkowski2001}. Limited by the count rate, we are unable to distinguish between these two NEI models statistically with current data. The fitting result generally supports a hot plasma with a temperature $kT\sim1$\keV\ and an ionization timescale $n_et\sim10^{10}$\,s\pcc\ for both two regions. The abundance of Si is over twice the solar abundance \citep{Wilms2000} for the best-fit model of the Shell, while that of O and Mg are near solar. For the Clump, the abundance seems to be similar but cannot be well constrained. A model describing plasma in CIE is also tested for these two regions. The best-fit CIE model gives similar fitting statistics to the NEI models but needs a very high metal abundance, especially for O and Mg, which often appears in the metal-rich ejecta of young or middle-aged SNRs \citep[see][and references therein]{Vink2020}. As O lines (0.5--0.7\keV) are dominated by astrophysical background compared to Mg ($\sim1.3$\keV) and Si ($\sim1.8$\keV) lines (see spectra in Appendix~\ref{app: spec}), we also fix O to solar abundance in CIE, named \vap* in Table~\ref{tab: spec}, and use F-test to test the significance of the under-ionization state (i.e. \vnei\ versus \vap). We noticed that a deviation from CIE is significant for the Shell with a null hypothesis probability of $4\times10^{-4}$, while for the Clump this improvement is less evident (probability $\sim0.017$). Therefore, we are in favor of the NEI model here but the CIE model cannot be completely excluded from fitting statistics as the ionization timescale and metal abundances need further constraints in the future study\footnote{However, the temperature of the CIE model is too high for a wind origin while it is hard to reach CIE for an SNR (see \S\ref{sec: ori} and Appendix~\ref{app: wind}\&\ref{app: sedov}).}. Meanwhile, as the diffuse emission has a low surface brightness, we also consider the ``null hypothesis'' where the emission is dominated by background fluctuation (i.e. fix the sky background and free its normalization instead of weighting by the sky area) as listed in Table~\ref{tab: spec}. This along with further analysis on the lower limit of the {\it norm} parameter indicates a robust detection of the diffuse emission (see Appendix~\ref{app: sig} for details).

\section{Discussion}
\subsection{Origin of the X-ray shell}
\label{sec: ori}
The newly discovered X-ray shell north of \ulx\ is not only much dimmer than the X-ray lobes but also composed of hot and under-ionized plasma, while the thermal emission of the lobe is much cooler ($<0.3$\keV) and close to CIE \citep{Safi-Harb2022}. Thus, we tend to exclude the possibility that the X-ray shell was heated by the past relativistic jet with an extreme precession angle up to $\sim90^{\circ}$, as we would expect a cooler and CIE plasma as found in bipolar X-ray lobes \citep{Safi-Harb2022}. Moreover, the radio polarization measurement of this region reveals tangential magnetic fields in the northern shell and a high polarization degree up to $\sim 50\%$ \citep{Gao2011, Farnes2017}, while magnetic fields in the jet region are found to align parallel to the jet direction \citep{Kaaret2023}. As observed in several old SNRs, highly ordered magnetic fields along the shell can be produced by a compression of the ambient gas by shocks \citep{Gao2011, Dubner2015}. Considering the spectroscopy and polarization properties, we suggest that the X-ray shell has a different physical origin from the elongated jet-blown lobes, but shows properties consistent with an SNR.

Based on the spectral fittings of the X-ray shell with an NEI model, we estimate its average density to be $\sim0.03$\pcc. This is lower than that of the terminal shock of the eastern lobe \citep[$\sim1$\pcc,][]{Safi-Harb2022}, showing that the gas density in the direction along and perpendicular to the jets is different.
The high temperature, low density, and faint optical emission \citep{Boumis2007, Farnes2017} here suggest that the remnant is still in a Sedov-Taylor phase \citep{Taylor1950, Sedov1959}. Under this assumption, we deduce that its age is $\sim17$--29\kyr\ and the explosion energy is $\sim0.4$--$1.4\times10^{51}$\erg\ (see details in Appendix~\ref{app: sedov}), a typical value for a core-collapse SN. The super-solar abundance of Si has previously been found in the outflowing matter of \ulx\ \citep{Brinkmann2005}, further supporting the existence of chemical enrichment near \ulx\ by an SNR.

\begin{figure*}
    \centering
    \includegraphics[width=\textwidth]{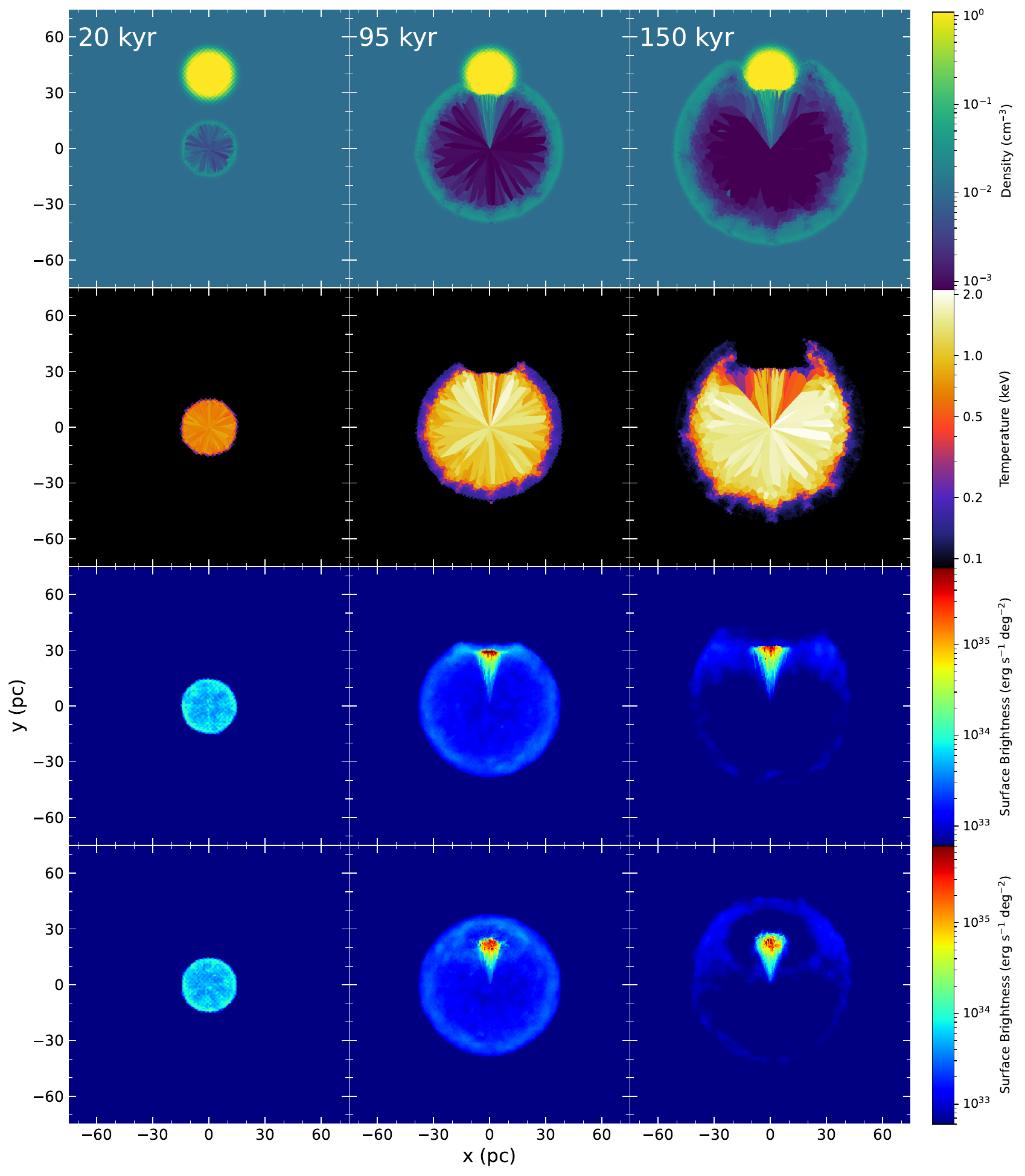}
    \caption{Evolution of a bubble powered by isotropic disk winds with a power of $3\times10^{37}$\erg\ps\ within a uniform medium and a dense cloud in the north. From top to bottom rows show the evolution of the density, temperature, and the X-ray surface brightness in 0.4--2.5\keV. The bottom panels give the surface brightness evolution for the case where the cloud is rotated 41.4$^\circ$  towards us (so that it is projected 30~pc north of \ulx).
    From left to right, the age of the wind bubble is 20, 95, and 150\kyr, respectively, corresponding to the age of the SNR, the beginning of the collision with the dense cloud, and the later evolution of the wind bubble.}
    \label{fig: wind}
\end{figure*}

Another potential physical origin of the X-ray shell could be the equatorial wind from the supercritical accretion disk \citep{Shakura1973, Churazov2024}, where fast winds have been observed with optical or radio observations \citep{Waisberg2019, Paragi1999}. We estimate the power of the equatorial wind as $\sim3\times10^{37}$\erg\ps, based on a mass loss rate of the gas flow of $10^{-4}$\ms\pyr\ \citep{Heuvel1981, Shkovskii1981, Fuchs2006} and an average wind terminal velocity of 1000\km\ps\ \citep{Fabrika1997}. This power is optimistically high since the wind velocity drops dramatically to $\sim100$\km\ps\ in areas perpendicular to the jet \citep{Fabrika2004}. Nevertheless, the wind power is much smaller than $\sim10^{39}$\erg\ps, the ``average power'' of a $\sim20$-kyr SNR. Based on a self-similar solution of wind bubble \citep{Weaver1977, MacLow1988}, the plasma would be much cooler ($\sim0.2$\keV) than the fitting result. To further assess this wind scenario, we simulate the evolution of an isotropic wind driven from \ulx\ into a uniform diffuse medium with a density of 0.01\pcc\ and also consider a dense cloud in a distance of 40\pc\ to mimic the dense-gas conditions near the Clump. The snapshot of the distribution of density, temperature, and estimated surface brightness at the age of 20, 95, and 150\kyr\ are shown in Figure~\ref{fig: wind}. For the estimation of the power of wind and supernova along with the details of the simulation, see Appendix~\ref{app: wind}.

Our simulation shows that it would take $\sim95$\kyr\ for the winds to blow out a cavity with a radius of the minor axis of W50. The challenge is that the wind is not strong enough for the observed X-ray flux and the gas temperature. The brightest emission appears where the winds interact with the cloud, but the surface brightness is still only half of the observed value of the Clump especially when projection effects are taken into consideration (see Appendix~\ref{app: wind} for more details). Without a wind-cloud interaction, other parts of the wind bubble are significantly fainter than the region Shell. Besides, we found that the temperature of the wind bubble is inconsistent with the observation. At around 95\kyr, the bubble shell has a temperature of $\lesssim 0.5$\keV\ (see Figure~\ref{fig: wind}), significantly cooler than that obtained from the observation. Theoretically, the temperature of the wind shell declines with the age due to the adiabatical cooling (see Equation \ref{eq: bubt}). The shell thus should be hotter at an earlier age, and the simulation gives an effective temperature (emissivity-weighted) of $\sim 0.6$\keV\ for the shell at 20\kyr.
At the wind-cloud interaction region where X-rays are enhanced, the effective temperature is less than 0.6\keV, either in the surface brightness peak ($\sim90$\kyr) or later evolution (150\kyr\ as an example). Moreover, a longer evolution timescale of the wind compared to the SNR scenario suggests likely a larger ionization timescale so that the plasma could be in or close to the CIE state.  Therefore, we conclude that the equatorial disk wind cannot describe the shell-like X-ray emission north of \ulx.

\subsection{Progenitor mass of the compact object in \texorpdfstring{\ulx}{}}
\label{sec: mass}
\begin{figure*}
    \centering
    \subfigure{\includegraphics[width=0.495\textwidth]{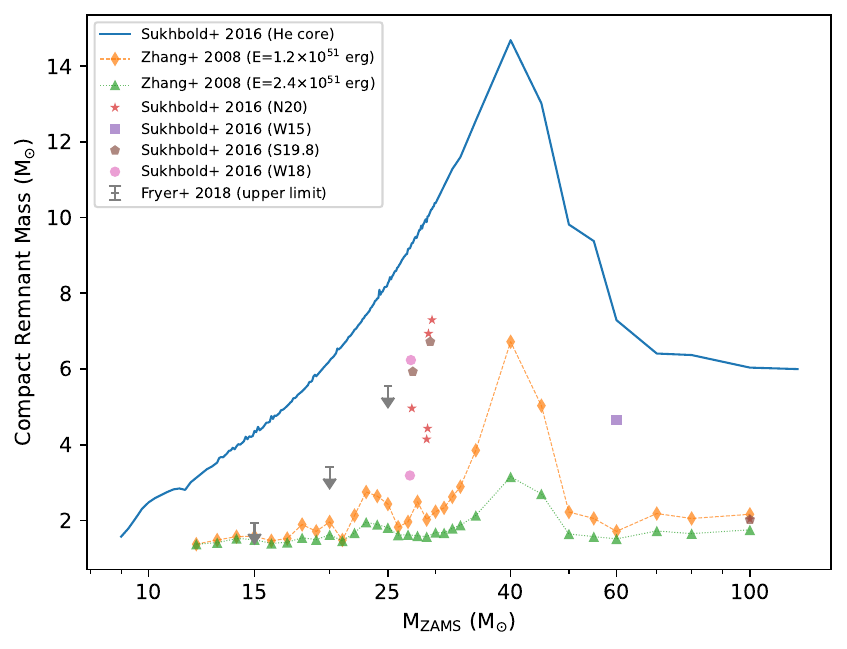}}
    \subfigure{\includegraphics[width=0.487\textwidth]{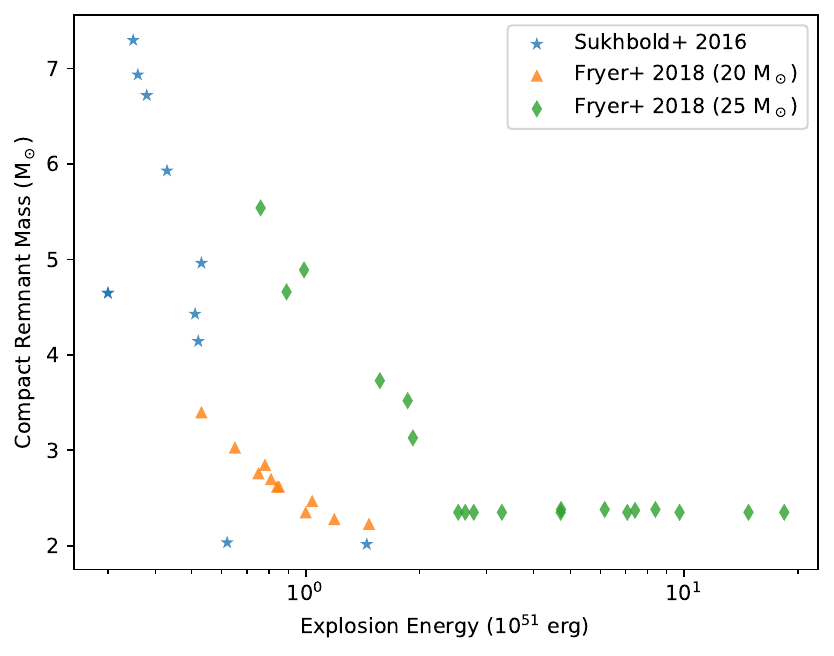}}
    \caption{{\it Left panel: }mass ``budget'' of the compact object. The blue solid line represents the mass of He core \citep{Sukhbold2016}. Yellow diamonds and green triangles correspond to models with a given explosion energy \citep{Zhang2008}. Grey scatters label the maximum compact remnant mass among different explosion parameters with a given progenitor mass \citep{Fryer2018}. The others are adopted from 4 different central engines \citep{Sukhbold2016} and only cases with a compact object mass over 2\ms\ are shown here. {Right panel: }mass of black holes with successful supernova explosion as a function of the explosion energy. Blue stars are given by \citet{Sukhbold2016}, while orange triangles (20\ms) and green diamonds (25\ms) are from \citet{Fryer2018}. Only cases where the black hole mass exceeds 2 \ms\ are reserved. }
    %for display.}
    \label{fig: rem}
\end{figure*}

As shown above, an SNR provides the best explanation for the newly discovered X-ray structure. Our X-ray study and the previously measured black hole mass of 5--15\ms\ in \ulx\ \citep{Cherepashchuk2019, Bowler2018, Gies2002} provide crucial information about the initial stellar mass and the SN properties for creating a black hole. Here we discuss the progenitor mass of the black hole in \ulx\ using the pre-SN core mass as predicted by the stellar evolution models \citep{Sukhbold2016, Woosley2015, Sukhbold2014, Woosley2007} along with core-collapse SN explosion models \citep{Zhang2008, Sukhbold2016, Fryer2018}. Figure~\ref{fig: rem} shows the pre-SN He core masses and the compact remnant mass as a function of the zero-age main-sequence stellar mass for stars with solar metallicity.

Previous studies suggest that the He core mass sets the upper limit of the compact object mass \citep{Zhang2008, Sukhbold2016}. For stars with initial mass ranging from 15 to 30\ms, we fitted the data given by \citet{Sukhbold2016} and found that there exists an approximate linear relation between stellar initial mass $M_\mathrm{ZAMS}$ and the final mass of He core $M_\mathrm{He}$: 
\begin{equation}
    M_\mathrm{ZAMS}\approx 2.44M_\mathrm{He}+4.75\,\mathrm{M_{\odot}}.
    \label{eq: mass}
\end{equation}

Above $\sim 40$\ms, the massive star strips most of the envelope mass with strong stellar winds and ends with an intermediate-mass core. To create a 5-M$_{\odot}$ compact object needs a progenitor star more massive than $\sim17$\ms, while a 10-M$_{\odot}$ compact object infers the progenitor mass range of $\sim30-50$\ms. Nevertheless, this estimate can be oversimplified as it does not consider the metal masses taken away by SN ejecta or the explosion parameters that significantly influence the final remnant mass.

Theoretical studies predict a wide diversity of SNe from these very massive stars, from failed SNe \citep[stars quietly collapse to black holes,][]{Kochanek2014} to ``hypernovae'' with $E_{\rm SN} \gtrsim10^{52}$\erg\ \citep{Nomoto2013}. To explore the dependence of the compact remnant mass on the explosion energy, we adopt the predicted results from three SN explosion models for stars with solar metallicities \citep{Zhang2008, Sukhbold2016, Fryer2018}. Two of these models are taken from \citet{Zhang2008} for explosion energies of  $\sim1.2\times10^{51}$\erg\ and $\sim2.4\times10^{51}$\erg. Four models from \citet{Sukhbold2016} applied different central engines (N20 \citep{Nomoto1988}, W15 \citep{Woosley2002}, S19.8 \citep{Woosley2002}, and W18 \citep{Sukhbold2016}) and provide a range of explosion energies (we only take those producing a remnant mass over 2.5\ms\ and with SN explosions). We also used models from \citet{Fryer2018}, which calculate the remnant masses and explosion energies for stars with initial masses of 15, 20, and 25\ms.

Figure~\ref{fig: rem}(a) illustrates the modeled remnant mass with the zero-age main-sequence stellar mass. The remnant masses from the SN models are much smaller than the He-core-confined mass, showing that a substantial amount of materials are ejected in the SN explosion. Despite some differences, all these explosion models found that $\gtrsim 5$\ms\ compact objects are made by stars $\gtrsim 25$\ms. As shown in Figure~\ref{fig: rem}(b), explosion energy strongly influences the compact remnant mass, with a heavier remnant favoring a less energetic explosion. 

Therefore, the normal SN explosion energy (0.7/$1.1\times 10^{51}$\erg) for W50 and the 5--15\ms\ compact object in \ulx\ \citep{Fabrika2004, Goodall2011, Farnes2017} suggest that this binary system contained a progenitor star with a mass $\gtrsim 25$\ms. Noteworthy, the mass of the black hole predicted by SN explosion models does not exceed 8\ms, although the upper limit of the black hole mass can be increased for failed SN cases. These models would be challenged if the mass of \ulx\ is proven heavier than 8\ms. Hence, a more accurate mass measurement of the compact object in \ulx\ is of great value.

\subsection{Supercritical accretion fed by fallback disk}
\label{sec: acc}
The existence of an SNR in W50 provides a more strict constraint on the accretion timescale and an alternative origin of accreted materials. It is thought that supercritical mass transfer of \ulx\ occurs due to the accretion from the donor star. To maintain the supercritical accretion, the donor star must be highly evolved and overfill the Roche Lobe \citep{Fabrika2004} within 20--30\kyr, the age of the SNR. Because of the short timescale of the post-main-sequence evolution, these two stars in the binary system would share almost the same mass, which happens with a low probability \citep{Heuvel1981}.

Alternatively, the accreted material may not purely originate from the companion star, and the SN fallback materials should also be considered. It has already been proposed that some of the young ultraluminous X-ray sources may be black holes accreting from their fallback disks \citep{Li2003}, but this hypothesis is waiting for an observational test. Here, we consider that a portion of metal-rich ejecta falls back to \ulx\ and provides materials for long-term accretion \citep{Li2003, Han2020}, while similar phenomena are observed around some neutron stars but on a smaller scale (e.g., 4U~0142+61 \citep{Wang2006} and 1E~161348$-$5055 in SNR RCW~103 \citep{Li2007}). On the assumption of a uniform distribution of 1-M$_{\odot}$ ejecta, we estimate the accretion rate of the fallback disk to be \citep{Mineshige1997, Li2003}: 

{\footnotesize
\begin{equation}
    \Dot{M}\approx(5.1\times10^{-5}\,\mathrm{M_{\odot}\,yr^{-1}})\left(\frac{M_c}{8\,\mathrm{M_{\odot}}}\right)\left(\frac{M_{acc}}{1\,\mathrm{M_{\odot}}}\right)\left(\frac{\alpha}{0.01}\frac{t}{10^4\,\mathrm{yr}}\right)^{-1.35},
\end{equation}
}
where $M_c$ is the mass of the compact object, $M_{acc}$ is the mass of the fallback materials, $\alpha$ is the viscosity parameter, and $t$ is the age. This is consistent with the observed mass loss rate of $10^{-5}-10^{-4}$\ms\pyr\citep{Heuvel1981, Shkovskii1981, Fuchs2006}. This amount of fallback is possible for creating a black hole \citep{Sukhbold2016, Fryer2018}, but the exact mass of the fallback materials is still highly uncertain. Further supporting evidence is that the abundance of Ni exceeds the solar value by $\sim10$ and $\sim2$, respectively, in the jet and winds of \ulx\ \citep{Brinkmann2005, Medvedev2018}. The Ni enhancement can be naturally explained as nucleosynthesis products released in an SN explosion rather than originating from the donor star's envelope.

\subsection{Possible contributions of an SNR to \texorpdfstring{$\gamma$}{}-ray emission}
\label{sec: tev}
W50 has been suggested to be a potential Galactic PeVatron, accelerating particles to 100's of TeV energies \citep{1999ApJ...512..784S}. After decades of searches for gamma-rays from this system, it was finally detected recently in the GeV and high-energy TeV bands \citep{Li2020, Fang2020, HAWC2018, Cao2023, HESS2024}, which has been attributed to the particle acceleration in the jets or winds. The identification of an SNR shell provides us with a new insight into the cosmic ray acceleration processes occurring in the system. Generally, the $\gamma$-ray emission is extended along the luminous X-ray jet lobes, with the peak emission spatially correlated with non-thermal X-ray knots \citep[and Mac Intyre et al. in prep.]{HAWC2018, Safi-Harb2022, HESS2024}. It is also worth noticing that some portion of the TeV and GeV emission, despite being faint, appear beyond the jet precession cone of \ulx\ \citep{Li2020, Fang2020, HESS2024}. To the west of \ulx\ where our study is focused, the TeV $\gamma$-ray emission extends to the north in all three bands. The excess seems to have a shell-like morphology, especially above 10\tev, and shares a similar angular distance to \ulx\ with the newly discovered X-ray shell. Unfortunately, the significance of the TeV emission associated with the X-ray shell is  $\le2\sigma$ based on the current \hess\ data. Modeling of the TeV gamma-ray emission suggests a particle acceleration timescale of $1-30$\kyr\ \citep{HESS2024}, consistent with the SNR age. The interpretation of the recent HESS observation suggests the jets are decelerated to speeds of $\sim0.08c$ by a shock at $\sim21'$ (30\pc\ at the distance of 5\kpc) away from \ulx\ \citep{HESS2024}, which would coincide nicely with the location of the X-ray shell. Given that SNRs are in general known to be powerful particle accelerators, it will be crucial to examine the role of the SNR in $\gamma$-ray emission in W50 and study the possible interaction between the SNR and jets \citep{Safi-Harb2022}. 

\begin{acknowledgments}
We appreciate the valuable comments and suggestions of the anonymous referee. This study is based on observations obtained with \xmm, an ESA science mission with instruments and contributions directly funded by ESA Member States and NASA. This publication utilizes data from Galactic ALFA HI (GALFA HI) survey data set obtained with the Arecibo L-band Feed Array (ALFA) on the Arecibo 305m telescope. The Arecibo Observatory is operated by SRI International under a cooperative agreement with the National Science Foundation (AST-1100968), and in alliance with Ana G. Méndez-Universidad Metropolitana, and the Universities Space Research Association. The GALFA HI surveys have been funded by the NSF through grants to Columbia University, the University of Wisconsin, and the University of California. We also acknowledge the use of archival X-ray data downloaded from HEASARC maintained by the NASA Goddard Space Flight Center.

Y.-H. C. acknowledges the foundation from the National Natural Science Foundation of China (NSFC) via grant NSFC 123B1021. The work of P. Z. is supported by grant NSFC 12273010. 
H. F. acknowledges funding support from the National Natural Science Foundation of China under grants Nos.\ 12025301, 12103027, \& 11821303, and the Strategic Priority Research Program of the Chinese Academy of
Sciences. X.-D. L. acknowledges support from the National Key Research and Development Program of China (2021YFA0718500) and the Natural Science Foundation of China under grant Nos. 12041301 and 12121003. 
S. M. and P. Z. also acknowledge support from Dutch Research Council (NWO) WARP program grant nr. 648.003.002 and S.M. acknowledges support from NWO VICI award nr. 639.043.513 and European Research Council (ERC) Synergy Grant ``BlackHolistic'' award nr. 101071643. 
S.S.H. acknowledges the support of NSERC (Natural Sciences and Engineering Research Council of Canada) through the Canada Research Chairs and Discovery Grants programs, and of the Canadian Space Agency.
\end{acknowledgments}

\software{SAS \citep{sas},  
          XSPEC \citep{xspec}, 
          Arepo \citep{Springel2010}
          }

\appendix
\section{ROSAT PSPC image}
\label{app: rosat}
Current \xmm\ data only cover a portion of the nebula W50 and thus fail to give us an overall view. Instead, we applied archival \rosat\ \citep{Truemper1982} PSPC data (US400271P, PI: M. Francis; WG500058P; PI: W. Brinkmann; see \citep{Safi-Harb1997}) for reference in favor of its large field of view. We mosaic the PSPC images in a scale of 30'' per pixel with the CIAO \citep{Fruscione2006} script {\it reproject\_image\_grid}. To increase the signal-to-noise ratio, we smoothed the PSPC image with a Gaussian kernel with a radius of 3 pixels. 

\section{Spectra with folded models}
\label{app: spec}

Here are the \xmm\ spectra with folded plasma models. Figure~\ref{fig: spec}(a) is the spectra of the background. Spectra of Shell are listed in Figure~\ref{fig: spec}(b)(c)(d) with \vap, \vnei, and \vps\ models, respectively. Figure~\ref{fig: spec}(e)(f) shows the spectra of Clump with \vap\ and \vnei\ models.

\begin{figure}
    \centering
    \subfigure{\includegraphics[width=0.496\textwidth]{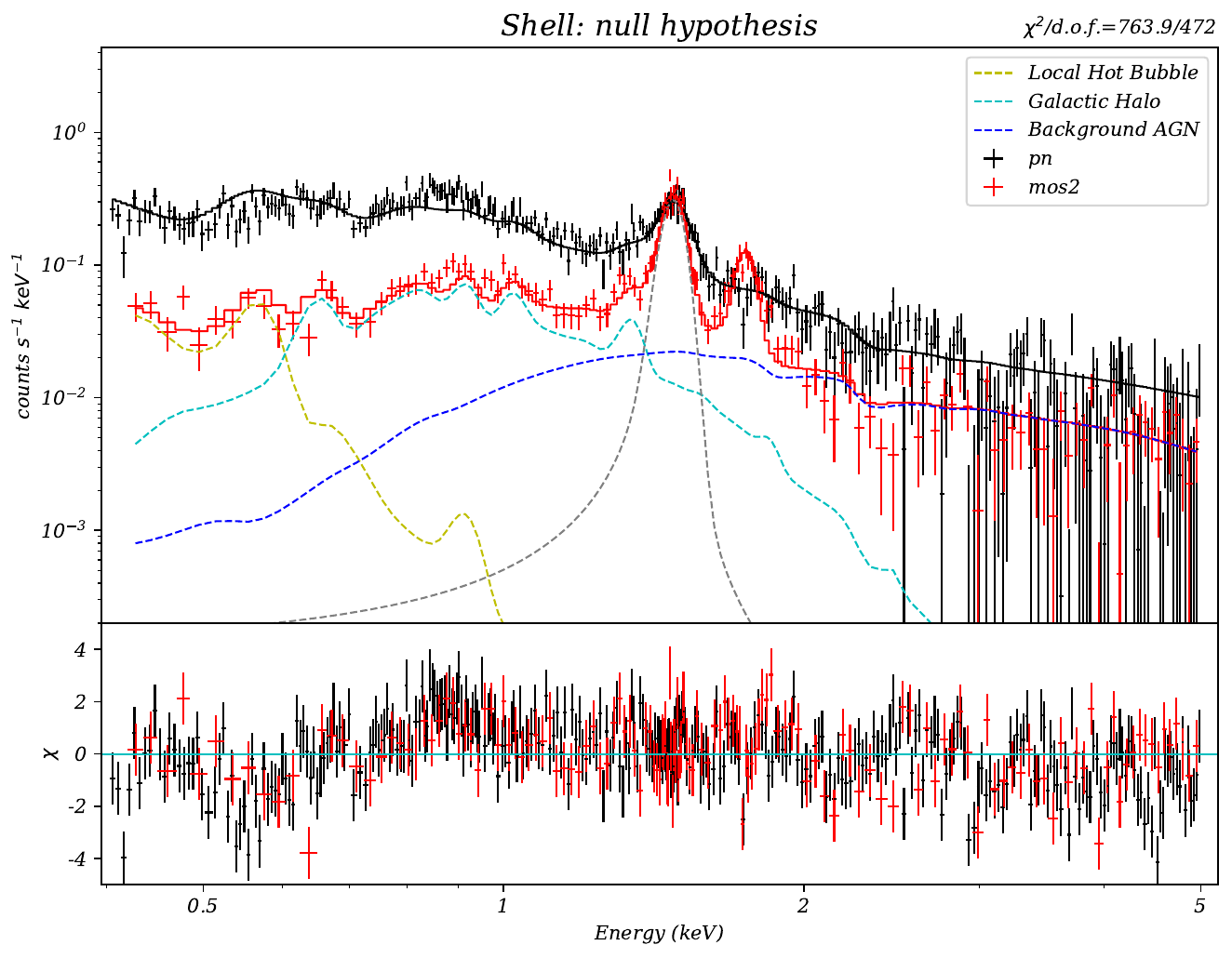}}
    \subfigure{\includegraphics[width=0.496\textwidth]{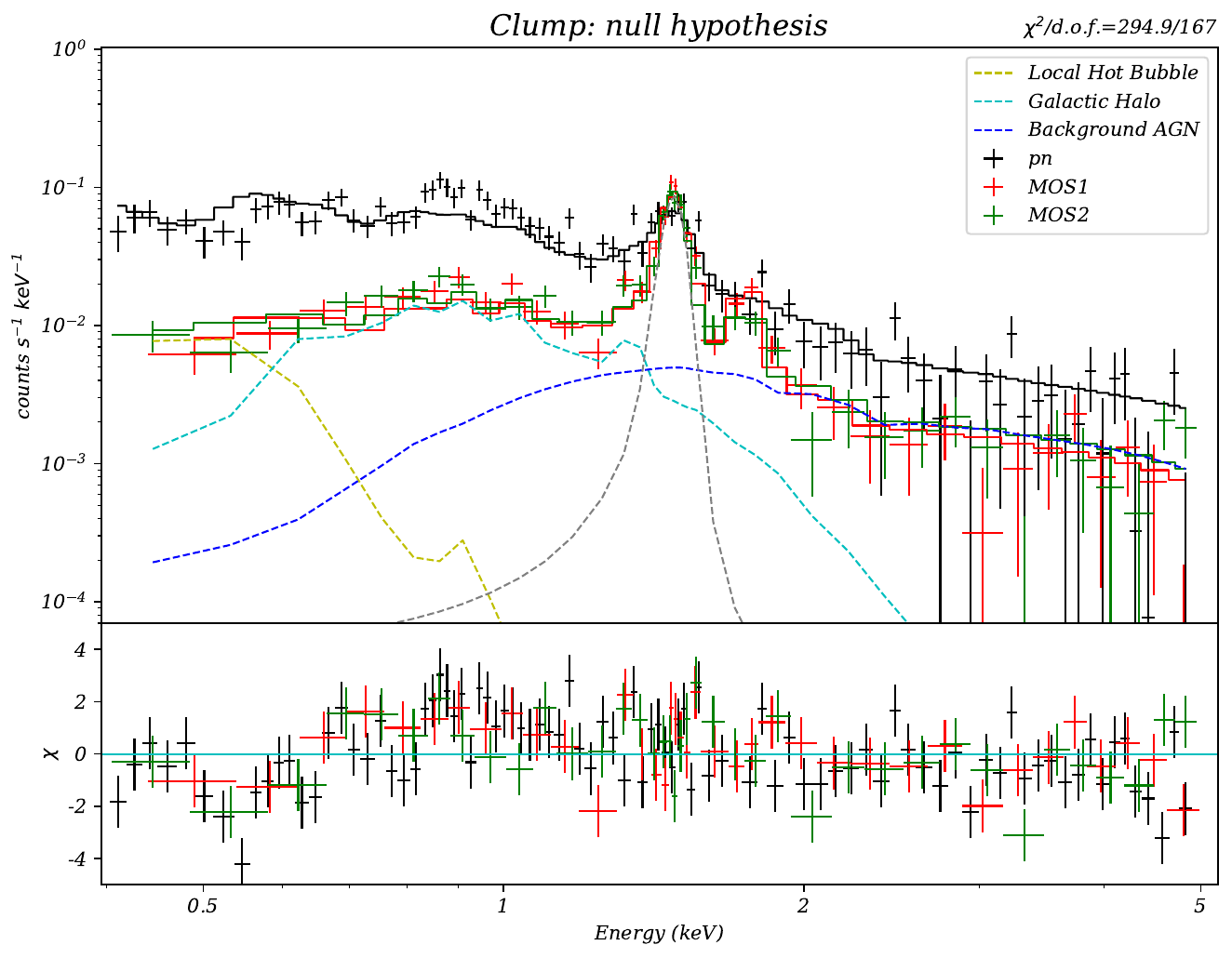}}
    \subfigure{\includegraphics[width=0.496\textwidth]{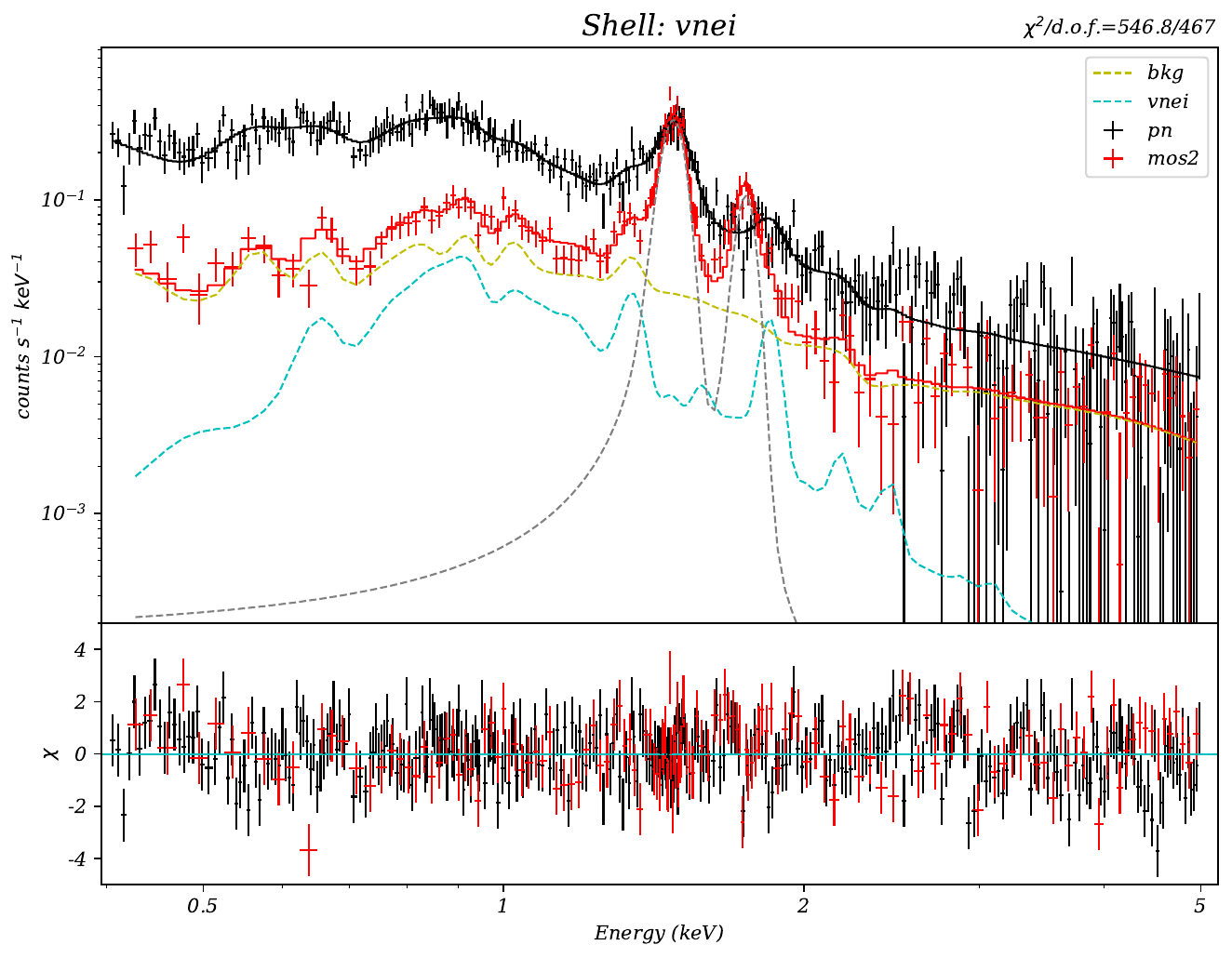}}
    \subfigure{\includegraphics[width=0.496\textwidth]{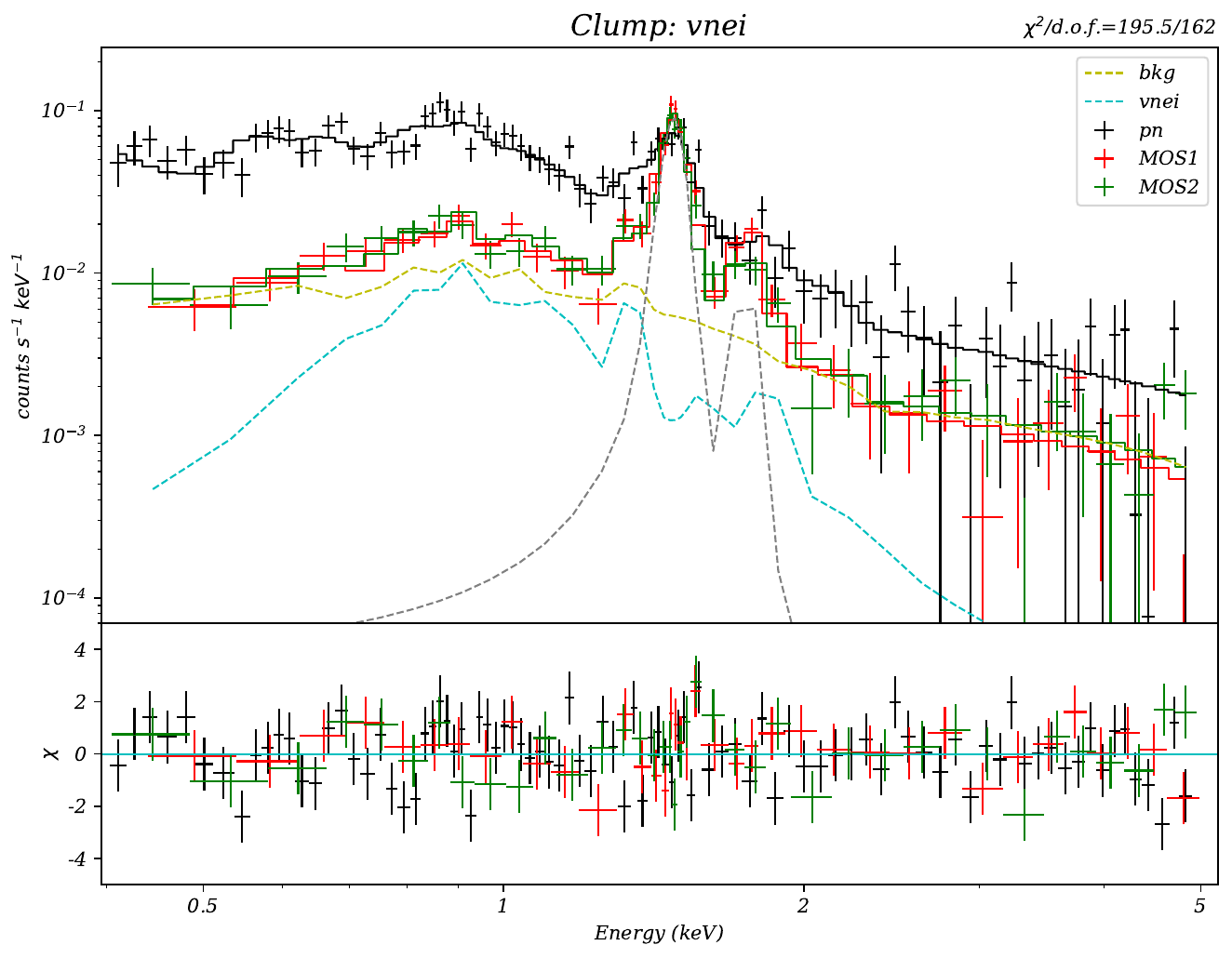}}
    \caption{Spectra with folded models of region Shell (left column) and Clump (right column). From top to bottom are null hypothesis, \vnei, \vps, \vap, and \vap* models. Different components of the MOS2 spectra are labeled with dashed lines. The null hypothesis spectra mark the astrophysical background from the local hot bubble, galactic halo, and background in yellow, cyan, and blue, respectively. For the others, the astrophysical background is labeled in yellow while the extra thermal component is marked in cyan. Grey lines represent the Al-K fluorescence lines of the instrument background.}
    \label{fig: spec}
\end{figure}

\begin{figure}
    \addtocounter{figure}{-1}
    \centering
    \subfigure{\includegraphics[width=0.496\textwidth]{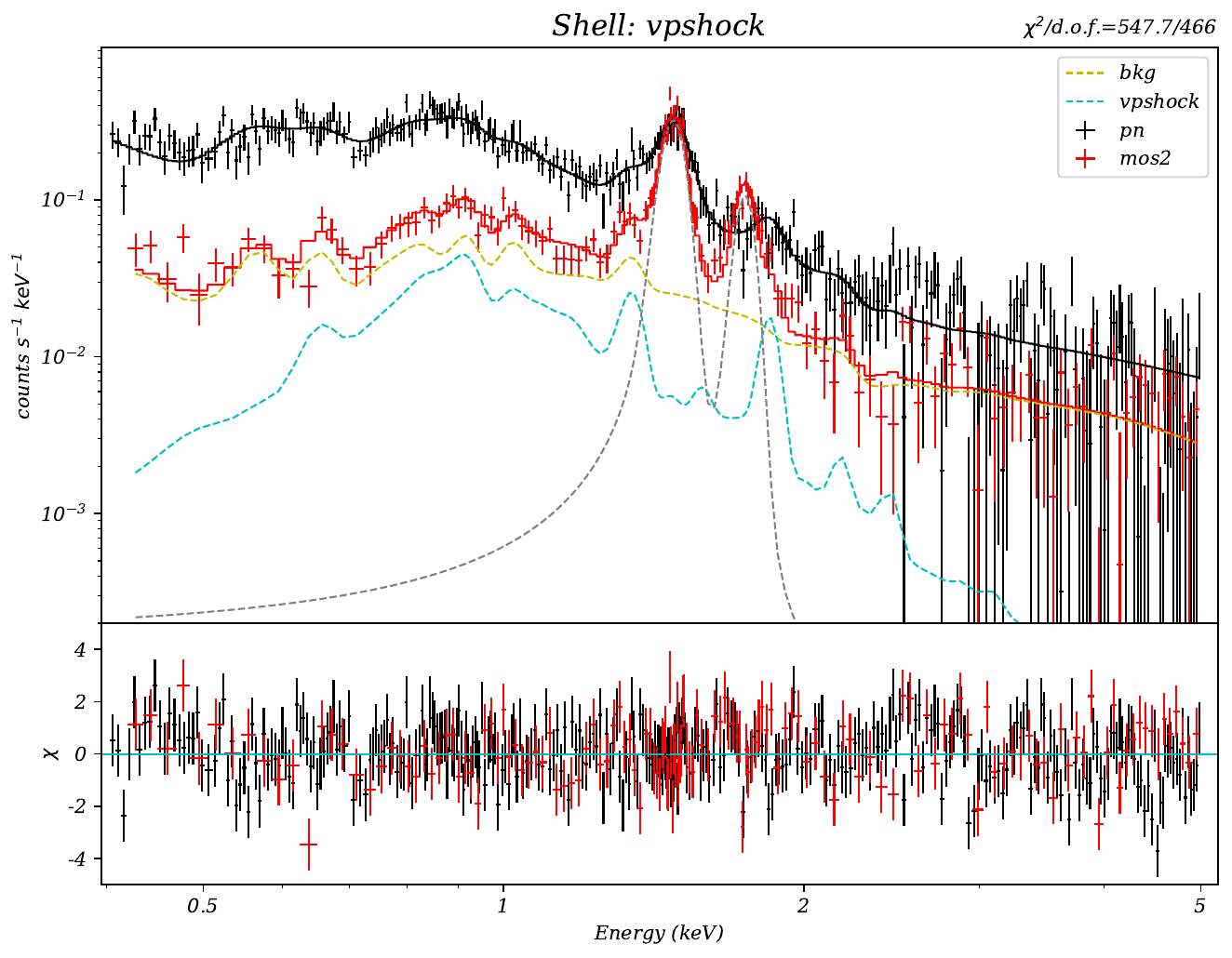}}
    \subfigure{\includegraphics[width=0.496\textwidth]{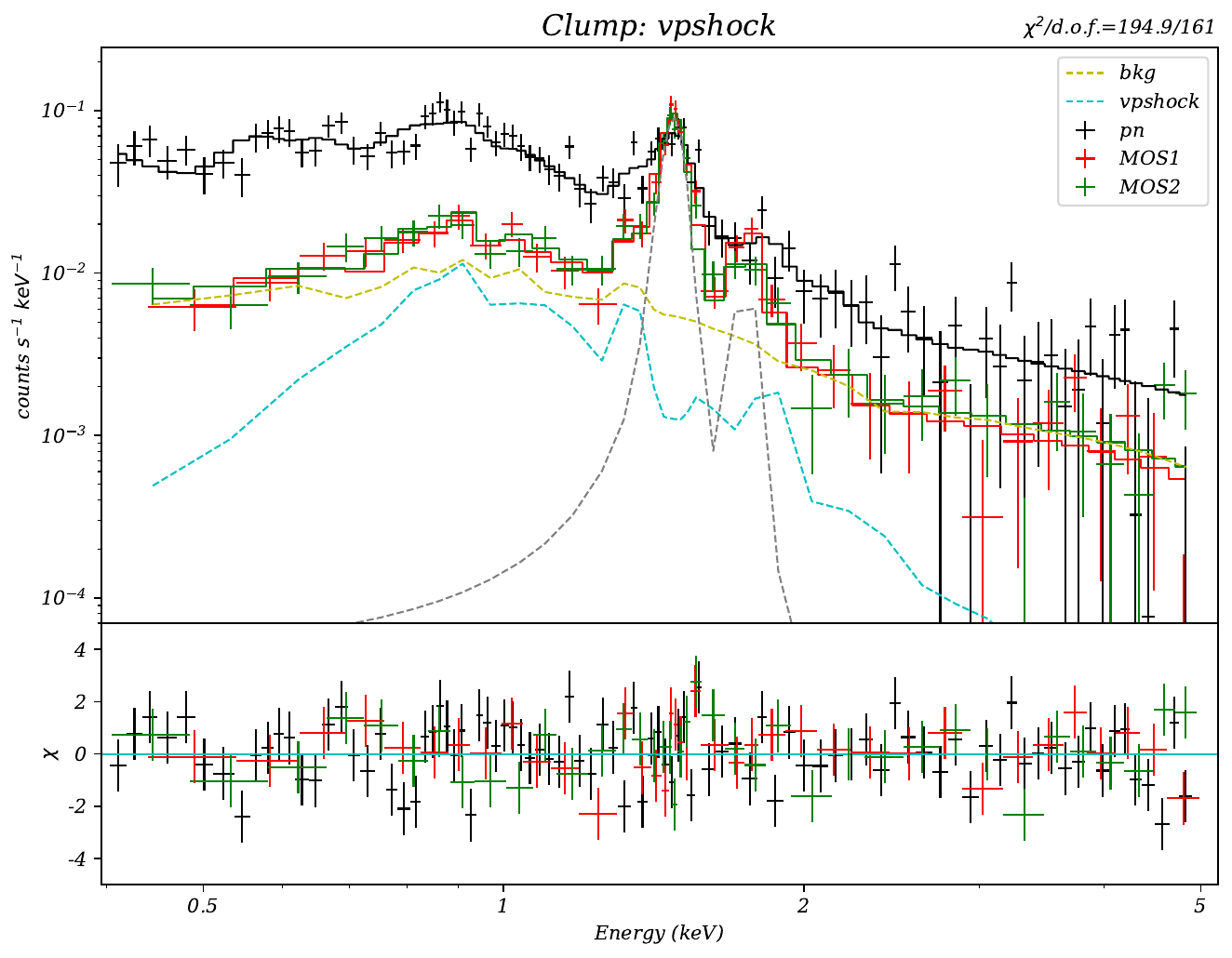}}
    \subfigure{\includegraphics[width=0.496\textwidth]{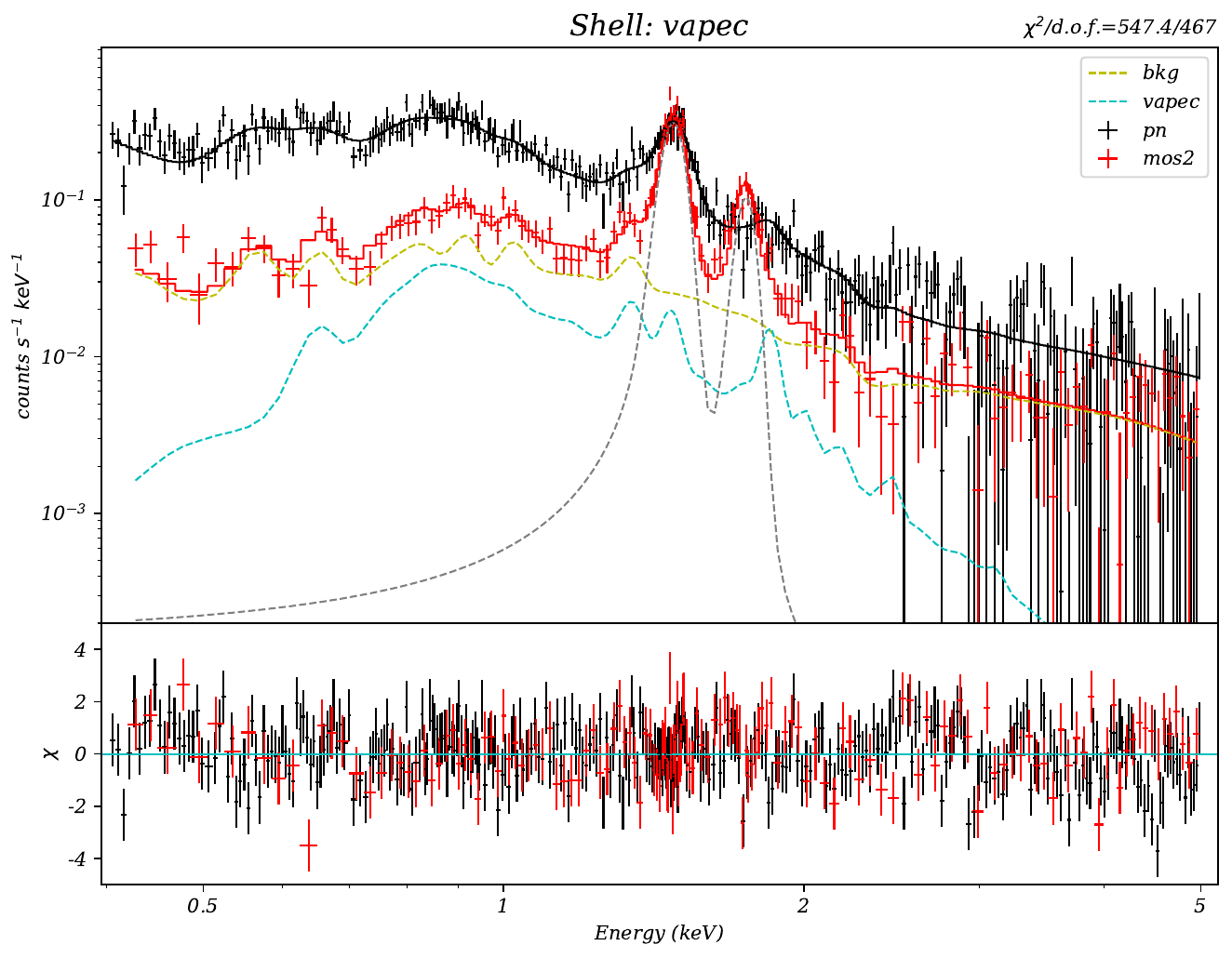}}
    \subfigure{\includegraphics[width=0.496\textwidth]{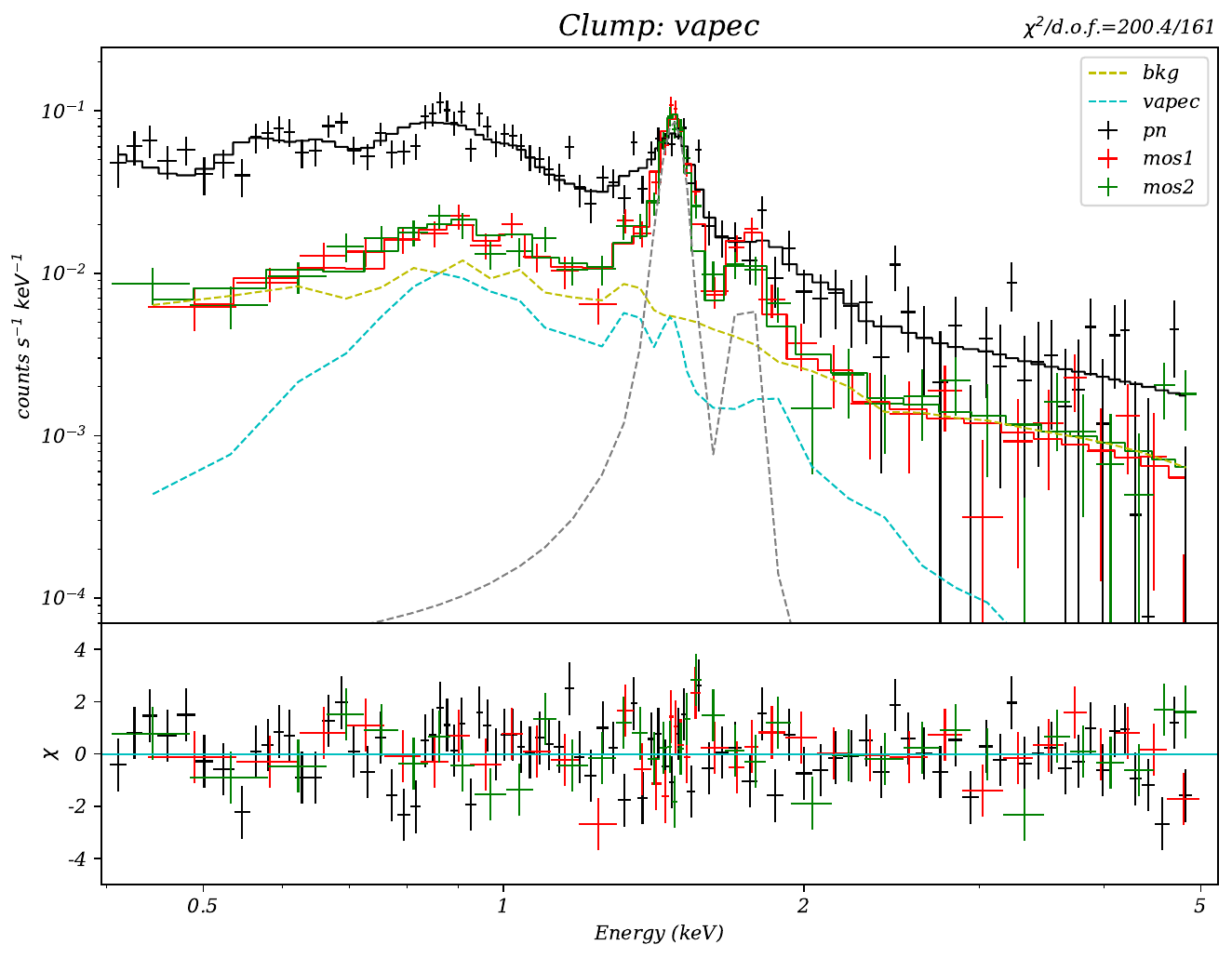}}
    \subfigure{\includegraphics[width=0.496\textwidth]{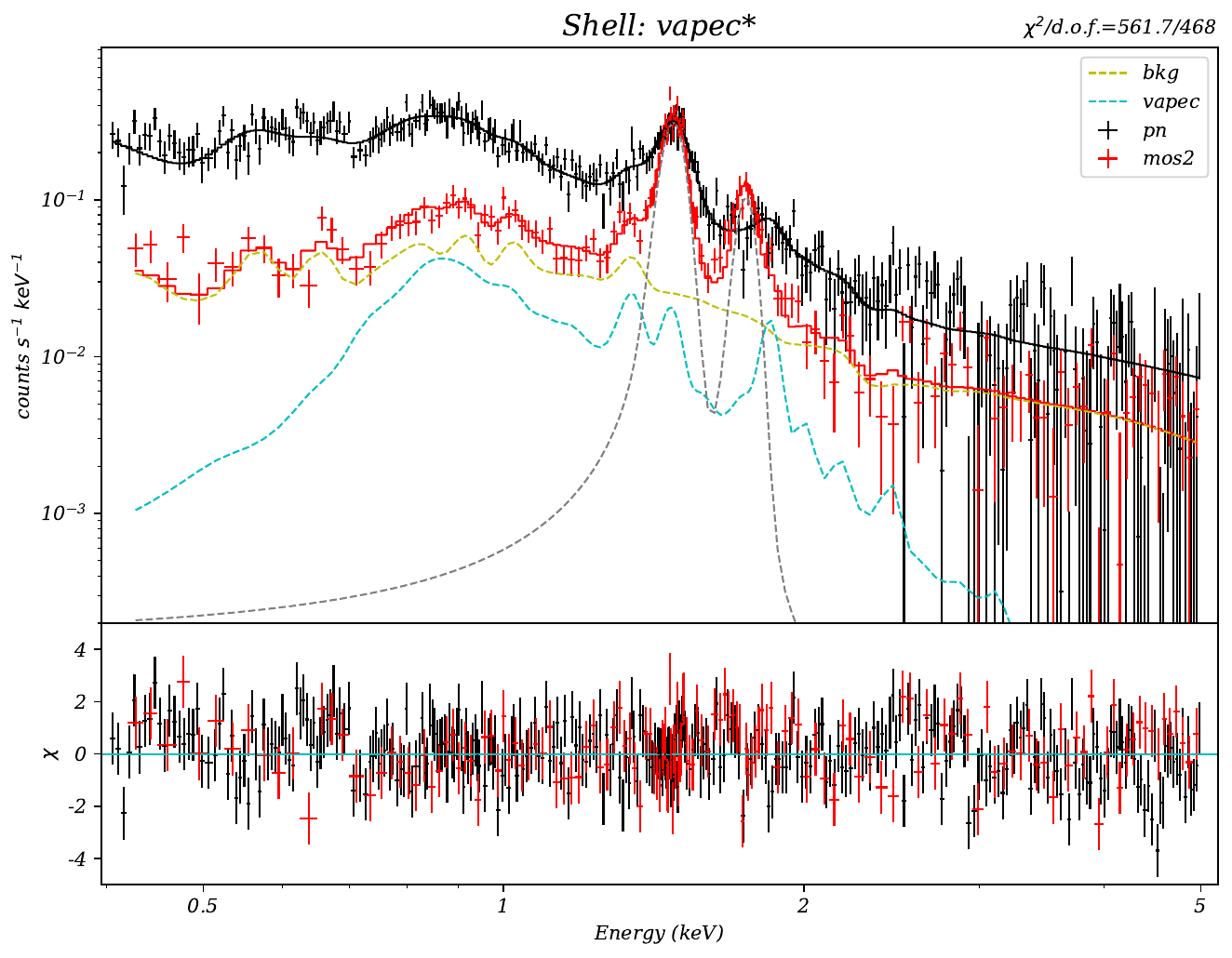}}
    \subfigure{\includegraphics[width=0.496\textwidth]{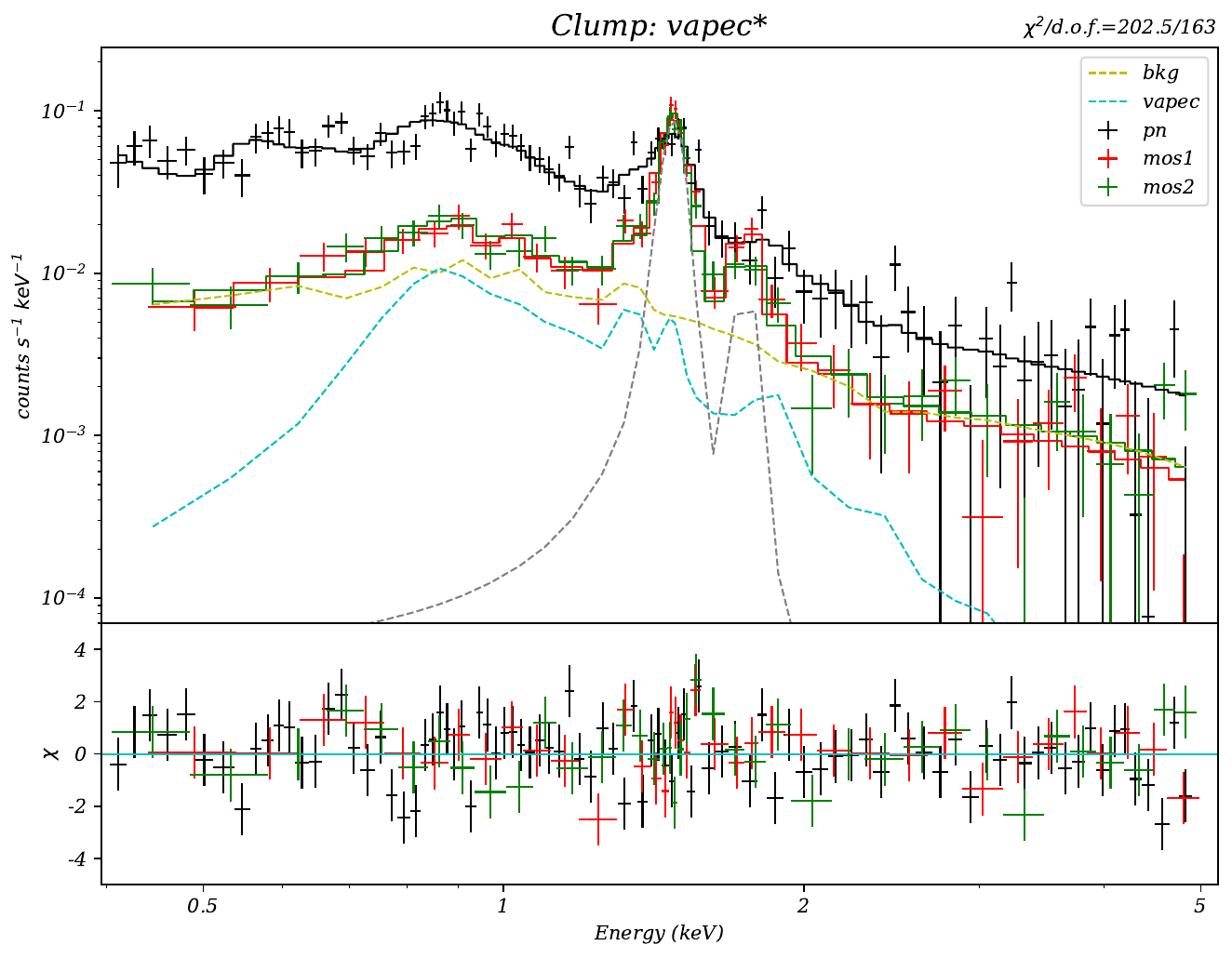}}    
    \caption{Continued.}
\end{figure}

\section{Significance of diffuse emission}
\label{app: sig}
Here we verify the significance of diffuse emission quantitatively mainly based on the following two methods. Firstly, the F-test is widely used to tell the significance of an extra component in spectral fittings \footnote{It should be noted that the F-test may be misleading \citep{Protassov2002}, but it still has a qualitative meaning here as the detection is evident.}. It would give a probability based on the chi-square and degrees of freedom of two different models, one of which has an extra component. A smaller probability (e.g. lower than $10^{-2}$ or $10^{-3}$) indicates higher significance. Besides different models describing the physical property of the diffuse emission mentioned in \S\ref{sec: result}, we also test the ``null hypothesis'' for comparison. Assuming that the diffuse emission originates from the fluctuation, we directly use the background model to fit the source spectra, and the results are also listed in Table~\ref{tab: spec}. Comparing the null hypothesis with the \vnei\ model, the probabilities given by the F-test are $5\times10^{-32}$ and $4\times10^{-13}$, for Shell and Clump, respectively.

\begin{figure}
    \centering
    \includegraphics[width=\textwidth]{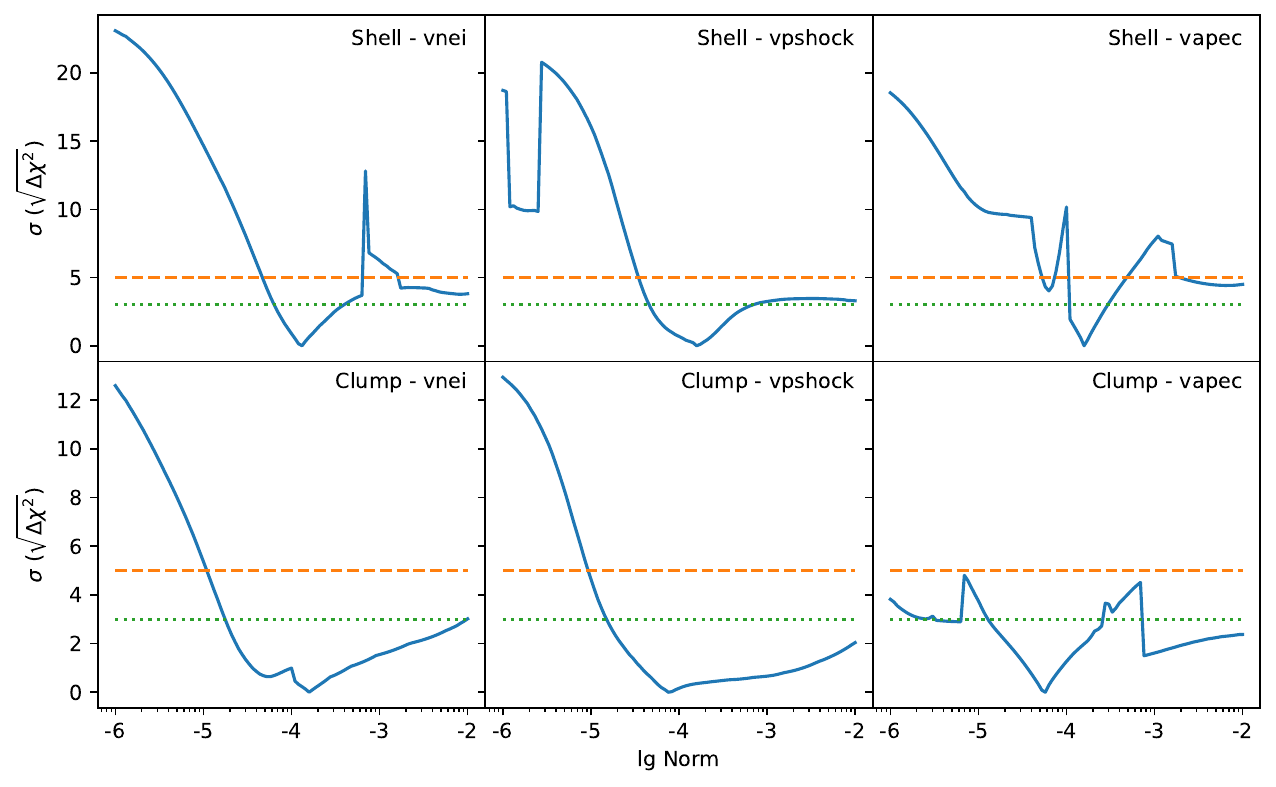}
    \caption{Relation between the change in the chi-square ($\Delta\chi^2$) and the value of emission measure (in units of cm$^{-5}$) given by {\it steppar} for three models in two regions (blue solid lines). Green dotted lines and orange dashed lines mark the fitting statistics corresponding to 3$\sigma$ ($\Delta\chi^2$=9) and 5$\sigma$ ($\Delta\chi^2$=25) confidence level. With large degrees of freedom, the chi-square distribution can be approximated as the Gaussian distribution.}
    \label{fig: step}
\end{figure}

Meanwhile, we study the probability distribution of the emission measure, or rather the parameter {\it norm} in Table~\ref{tab: spec}, which is proportional to the unabsorbed flux. As the parameter error is asymmetric and non-Gaussian, we use the XSPEC script {\it steppar} to fit the spectra while stepping the value of the emission measure logarithmically from $10^{-6}\,\mathrm{cm}^{-5}$ to $10^{-2}\,\mathrm{cm}^{-5}$. The chi-square increases rapidly when the emission measure drops below $\sim10^{-5}\,\mathrm{cm}^{-5}$ and have a 5$\sigma$ lower limit around this value as plotted in Figure~\ref{fig: step}. Therefore, we conclude that the detection of the diffuse emission is robust.

\section{Supernova remnant in Sedov-Taylor phase}
\label{app: sedov}
The physical properties of the Shell are calculated using the fitting results of NEI models. The density depends on the volume of X-ray-emitting gas and here we assume the Shell and Clump have an average depth of $\sim5'$ (approximately the width of the Shell region) at a distance of 5\kpc\ \citep{Su2018}. The region area is based on the {\it backscal} keyword of the spectra. Hence, we estimate the average density of the Shell is: 
\begin{equation}
    n=\sqrt{\frac{4\times10^{14}\pi\,norm\,D^2}{1.2\eta V}}
        \approx(0.03\,\mathrm{cm^{-3}})\left(\frac{norm}{1.4\times10^{-4}\,\mathrm{cm^{-5}}}\right)^{\frac{1}{2}}\left(\frac{D}{5\,\mathrm{kpc}}\right)^{-\frac{1}{2}},
\end{equation}
where $D\sim5$\kpc\ is the distance of W50 \citep{Su2018}. For the Clump where X-ray emission is enhanced, its density is a little higher with $n\approx(0.05\,\mathrm{cm^{-3}})(norm/10^{-4}\,\mathrm{cm^{-5}})^{1/2}(D/5\,\mathrm{kpc})^{-1/2}$.

Noticeably, this density is much lower than that in the X-ray lobe of $\sim$1\pcc\ 95\pc\ away from \ulx\ \citep{Brinkman2007, Safi-Harb2022}, 
indicative of different gas densities in the two regions. Assuming the X-ray shell is located at the radio boundary of W50 ($28'$ to \ulx) but projected inside the nebula, the distance of the Shell to \ulx\ is obtained as: 
\begin{equation}
    R=(41\,\mathrm{pc})\,\left(\frac{D}{5\,\mathrm{kpc}}\right)
\end{equation}
and the gas mass swept by the shock $M_s$ (assumed in a uniform medium) is: 
\begin{equation}
    M_s=\frac{4}{3}\pi R^3\ 1.4m_pn_0=(74\,\mathrm{M_{\odot}})\left(\frac{norm}{1.4\times10^{-4}\,\mathrm{cm^{-5}}}\right)^{\frac{1}{2}}\left(\frac{D}{5\,\mathrm{kpc}}\right)^{\frac{5}{2}},
\end{equation}
where $m_p$ is the proton mass and $n_0=n/4$ is the density of pre-shock gas. As the swept mass exceeds that of the ejecta (in general $<10$\ms\ \citep{Sukhbold2016}) the SNR should have left the free expansion phase and entered the Sedov phase \citep{Sedov1959, Taylor1950} when the SNR evolves adiabatically. The shock velocity is also typical for SNRs in the Sedov-Taylor phase, but larger than that in the radiative phase \citep[$\lesssim200$\km\ps,][]{Vink2020}.

With the temperature of the plasma, we can derive the shock velocity: 
\begin{equation}
\label{eq: shock}
    v_s=\sqrt{\frac{16kT}{3\mu m_p}}=(923\,\mathrm{km\,s^{-1}})\,\left(\frac{kT}{1\,\mathrm{keV}}\right)^{\frac{1}{2}}
\end{equation}
where $\mu=0.6$ is the mean atomic weight for fully ionized plasma. %per mass.
Assuming a uniform ambient ISM, the age of W50 is
\begin{equation}
    t=\frac{2R}{5v_s}=(17.4\,\mathrm{kyr})\,\left(\frac{D}{5\,\mathrm{kpc}}\right)\left(\frac{kT}{1\,\mathrm{keV}}\right)^{-\frac{1}{2}}.
    \label{eq: age}
\end{equation}

With age in Equation~\ref{eq: age}, we can calculate the expected ionization timescale of the hot gas and compare it with the X-ray spectral fitting results: 
\begin{equation}
    \tau=n_et=(2.0\times10^{10}\,\mathrm{s\,cm^{-3}})\left(\frac{norm}{1.4\times10^{-4}\,\mathrm{cm^{-5}}}\right)^{\frac{1}{2}}\left(\frac{D}{5\,\mathrm{kpc}}\right)^{\frac{1}{2}}\left(\frac{kT}{1\,\mathrm{keV}}\right)^{-\frac{1}{2}},
    \label{eq: tau}
\end{equation}
which is consistent with our spectral analysis ($10^{10}-10^{11}$\,s\pcc). Then we obtained the explosion energy of
\begin{equation}
    E_{\rm SN}=\frac{R^5}{\xi t^2}\ 1.4m_pn_0=(9.3\times10^{50}\,\mathrm{erg})\,\left(\frac{norm}{1.4\times10^{-4}\,\mathrm{cm^{-5}}}\right)^{\frac{1}{2}}\left(\frac{D}{5\,\mathrm{kpc}}\right)^{\frac{5}{2}}\left(\frac{kT}{1\,\mathrm{keV}}\right),
    \label{eq: exp}
\end{equation}
where dimension-less factor $\xi=2.025$. This is a typical value for a core-collapse SN \citep{Vink2020}.

As a high-mass X-ray binary, \ulx\ should have a massive progenitor system, which can drive strong stellar winds to create a large, low-density wind-blown bubble. If the SNR is evolving in a circumstellar medium with a density $n\propto r^{-2}$ (r is the distance to the central star), the age is derived as $t=\frac{2R}{3v_s}\approx28.6$\kyr\ \citep{Vink2020}, higher than in uniform medium but in the same order. 
In the wind bubble scenario, the dimension-less factor in Equation~\ref{eq: exp} can be expressed approximately by \citet{Truelove1999}: 
\begin{equation}
    \xi=\frac{(5-s)(10-3s)}{8\pi},
\end{equation}
where $s=2$ for a wind bubble. Applying it along with the age of 28.6\kyr\ as mentioned above to the Equation~\ref{eq: exp}, the explosion energy would be $\sim1.4\times10^{51}$\erg, $\sim1.5$ times that in a uniform environment. It should be noted that even adopting the fitting results of the CIE model shown in Table~\ref{tab: spec}, the age (Equation~\ref{eq: age}) and explosion energy (Equation~\ref{eq: exp}) do not change significantly. Moreover, the expected ionization timescale (Equation~\ref{eq: tau}) is still $\lesssim10^{11}$\,s\pcc, not reaching CIE yet.

Another factor that needs considering is that the heating timescale of electrons can be longer than the age of an SNR \citep[see][and references therein]{Borkowski2001, Ghavamian2013} but in Equation~\ref{eq: shock} the electrons and ions are assumed with the same temperature. Here we use the measured ionization timescale $\tau$, although with higher uncertainties compared to the temperature, to recalculate the age in Equation~\ref{eq: age}:
\begin{equation}
    t=\frac{\tau}{n_e}=(26.4\mathrm{\kyr})\,\left(\frac{\tau}{3\times10^{10}\,\mathrm{s\,cm^{-3}}}\right)\left(\frac{norm}{1.4\times10^{-4}\,\mathrm{cm^{-5}}}\right)^{-\frac{1}{2}}\left(\frac{D}{5\,\mathrm{kpc}}\right)^{\frac{1}{2}}
\end{equation}

and the explosion energy in Equation~\ref{eq: exp}: 
\begin{equation}
    E_{\mathrm{SN}}=(4.0\times10^{50}\mathrm{\erg})\,\left(\frac{norm}{1.4\times10^{-4}\,\mathrm{cm^{-5}}}\right)^{\frac{3}{2}}\left(\frac{D}{5\,\mathrm{kpc}}\right)^{\frac{7}{2}}\left(\frac{\tau}{3\times10^{10}\,\mathrm{s\,cm^{-3}}}\right)^{-2}.
\end{equation}
This explosion energy is slightly lower than that given in Equation~\ref{eq: exp} but is also reasonable for a core-collapse SNR.

It should be noted that the age and explosion energy here are based on strong assumptions and have uncertainty. Besides the fitting errors in Table~\ref{tab: spec}, the real spatial structure of X-ray-emitting gas is unknown. Meanwhile, the evolution model is based on isotropic explosion and uniform density distribution, while asymmetric explosion may happen for core-collapse SNRs \citep[e.g.,][]{Lopez2011}, and current data reveal the density gradients. Anyhow, the calculation here is still representative, indicating that a black hole could form with a canonical SN.

\section{Equatorial Disk Wind}
\label{app: wind}
The equatorial disk wind may also blow out a bubble. For \ulx, the disk wind velocity has been measured based on independent optical and radio observations for the regions with a polar angle $\alpha>$ 60°. The wind velocity is $\sim1300$\km\ps\ and $\sim100$\km\ps\ when $\alpha$ is 60° and 90° respectively, while an average value is $\sim450$\km\ps\ \citep{Fabrika2004, Waisberg2019}. The mass transfer rate as the function of polar angle is unclear, but the total mass loss rate is $10^{-4}$\ms\pyr\ \citep{Heuvel1981, Shkovskii1981, Fuchs2006} and it grows as the polar angle $\alpha$ increases \citep{Fabrika2004}. For simplicity, we assume the wind is isotropic with a mass loss rate of $\Dot{M}_w=10^{-4}$\ms\pyr\ and a wind velocity of $v_w=1000$\km\ps, which is around the upper limit of the average wind velocity. Assuming a constant wind, the power of the wind is: 
\begin{equation}
    L_w=\frac{1}{2}\Dot{M}_w v_w^2=3\times10^{37}\,\mathrm{erg\,s^{-1}}.
\end{equation}

To compare with the SNR scenario, we also calculate the ``average power'' of the SN by simply dividing the explosion energy by the age: 
\begin{equation}
    L_{SN}=\frac{E_{\rm SN}}{t}=(1.7\times10^{39}\,\mathrm{erg\,s^{-1}})\,\left(\frac{norm}{1.4\times10^{-4}\,\mathrm{cm^{-5}}}\right)^{\frac{1}{2}}\left(\frac{D}{5\,\mathrm{kpc}}\right)^{\frac{3}{2}}\left(\frac{kT}{1\,\mathrm{keV}}\right)^\frac{3}{2},
\end{equation}
which is significantly higher than the power of equatorial disk wind.

Thus, the disk wind is far from a powerful energy source. For a better illustration, we applied a toy evolution model of a wind bubble based on the self-similar solution \citep{Weaver1977, MacLow1988}. The radius of the bubble $R$ can be expressed as: 
\begin{equation}
    R=(10.6\,\mathrm{pc})\left(\frac{L_{37}t_5^3}{n_0}\right)^{1/5},
    \label{eq: radius}
\end{equation}
where $L_{37}$, $t_5$, and $n_0$ are the wind power, bubble age, and ambient density in units of $10^{37}$\erg\ps, $10^5$\yr, and 1\pcc, respectively. In a wind bubble, the temperature decreases from the center to the edge: 
\begin{equation}
    kT=(0.39\,\mathrm{keV})L_{37}^{8/35}n_0^{2/35}t_5^{-6/35}(1-x)^{2/5},
    \label{eq: bubt}
\end{equation}
where $x=\frac{r}{R}$ is the dimensionless distance as the ratio of the distance from the bubble center $r$ to the bubble radius $R$. Its mean value of the whole bubble is: 
\begin{equation}
    k\Bar{T}=(0.21\,\mathrm{keV})L_{37}^{8/35}n_0^{2/35}t_5^{-6/35}.
\end{equation}
With this equation and Equation~\ref{eq: radius}, we can get a relation among the bubble size, temperature, and wind power independent of its age and ambient gas: 
\begin{equation}
    k\Bar{T}=(0.19\,\mathrm{keV})\left(\frac{L_w}{3\times10^{37}\,\mathrm{erg\,s^{-1}}}\right)^{2/7}\left(\frac{R}{40\mathrm{\pc}}\right)^{-2/7}.
\end{equation}
Therefore, if the disk wind contributes to the diffuse X-ray emission, the temperature would be significantly lower than the 3$\sigma$ lower limit (0.71\keV\ for \vnei\ and 0.55\keV\ for \vps) according to spectral analysis of the northern X-ray shell.
The index $\frac{2}{7}$ suggests the temperature is not sensitive to the size of W50 and the wind power.

We use the moving-mesh hydrodynamical software Arepo \citep{Springel2010} to simulate the evolution of the wind-blown bubble and the collision between the isotropic disk wind and a dense cloud 40\pc\ north of \ulx. The simulation box size is 150\pc$\times$150\pc$\times$150\pc. We initially set $5.2\times10^5$ Voronoi cells, placing a 10-M$_{\odot}$ black hole particle in the center of the simulation box. The number density of the dense cloud is defined by a spherically symmetric Gaussian distribution. The maximum density is set to 120\pcc\ and its 3$\sigma$ radial size is $15'$, corresponding to a total mass of 25,000\ms\ given by \citet{Li2020}. Besides the initial dense cloud, we set a uniform density background of 0.01\pcc\ inferred from our X-ray spectral analysis. The dense cloud and the homogeneous diffuse gas have a uniform initial gas temperature of $10^4$\,K. 

In every time step of our simulation, we find 32 cells closest to the central black hole and add the corresponding mass, momentum, and kinetic energy of the disk wind to the cells, weighted by the cell solid angles extended to the central point. We include the self-gravity of the gas and multiple cooling mechanisms, which contain thermal bremsstrahlung, hydrogen and helium recombination, and solar-abundance metal line cooling \citep{Li2019}. However, we emphasize that neither self-gravity nor cooling plays an important role in our simulation. 

The simulation is applied until the age of the system reaches 224\kyr. We firstly let the cloud on the plane perpendicular to the line of sight to better display the number density $n$ and temperature $T$ distributions of the wind bubble in the simulation snapshots. Then, as the X-ray emission looks inside the radio shell (Figure~\ref{fig: 4band}) probably due to the projection effect, we also integrate the surface brightness in a direction where the perpendicular distance from the cloud to the central black hole is 30\pc. To estimate the X-ray emission intensity, we use {\it Xspec} to calculate the absorbed X-ray emissivity (unit: erg\ps\cm$^{-3}$) of plasma with a given temperature from 0.1 to 3.0\keV. The foreground column density is set to $7\times 10^{21}$\cm$^2$ based on the spectral analysis above. By integrating the emissivity along the line of sight, we can estimate the surface brightness (unit: erg\ps\psqd) of the bubble in 0.4--2.5\keV. The snapshots of the gas number density, gas temperature, and surface brightness map of our simulation are shown in Figure~\ref{fig: wind}.

\begin{figure}
    \centering
    \subfigure{\includegraphics[width=0.49\linewidth]{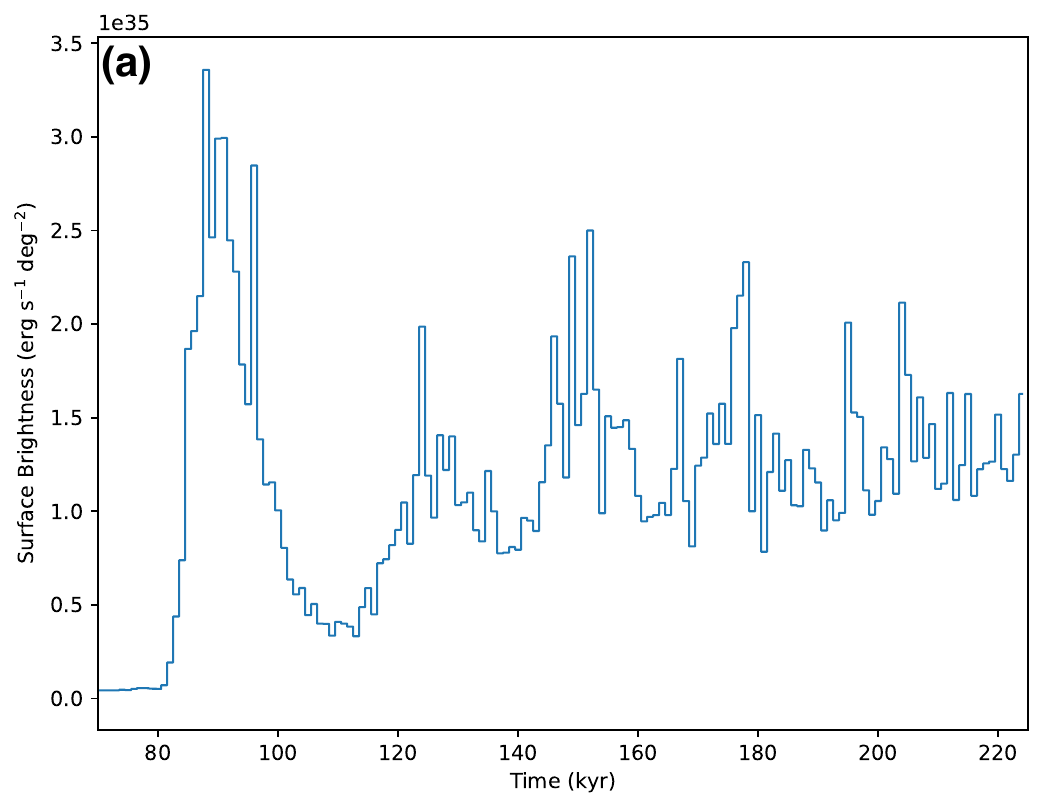}}
    \subfigure{\includegraphics[width=0.49\linewidth]{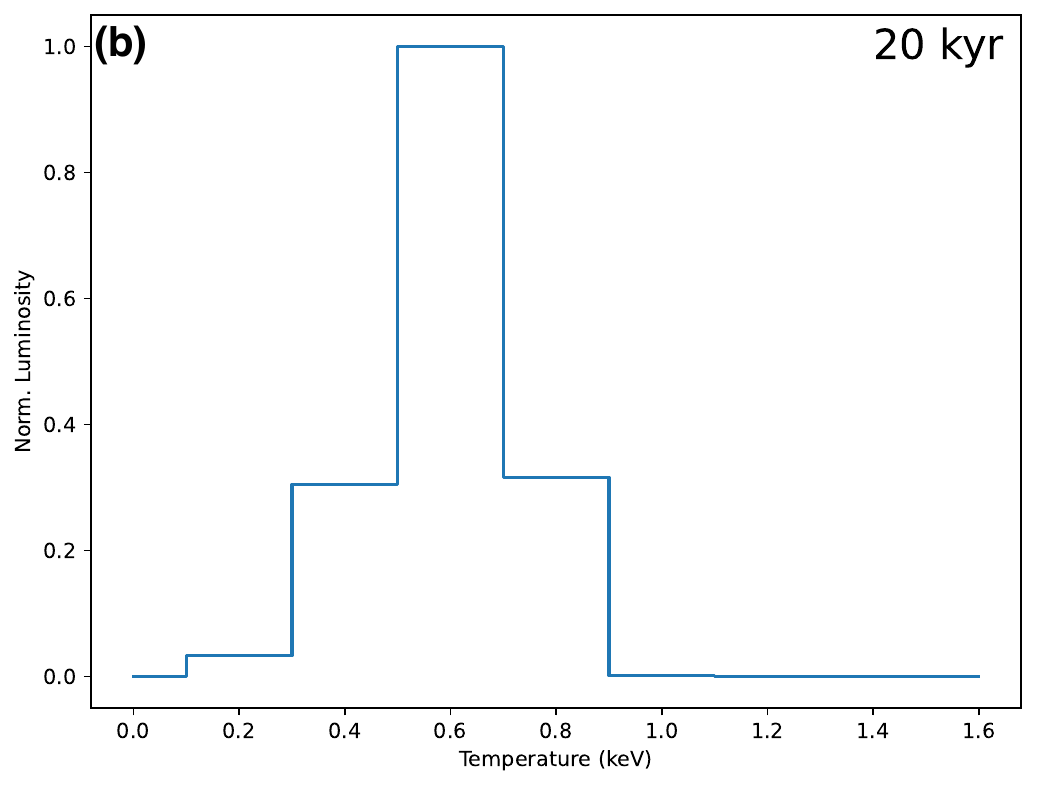}}
    \subfigure{\includegraphics[width=0.49\linewidth]{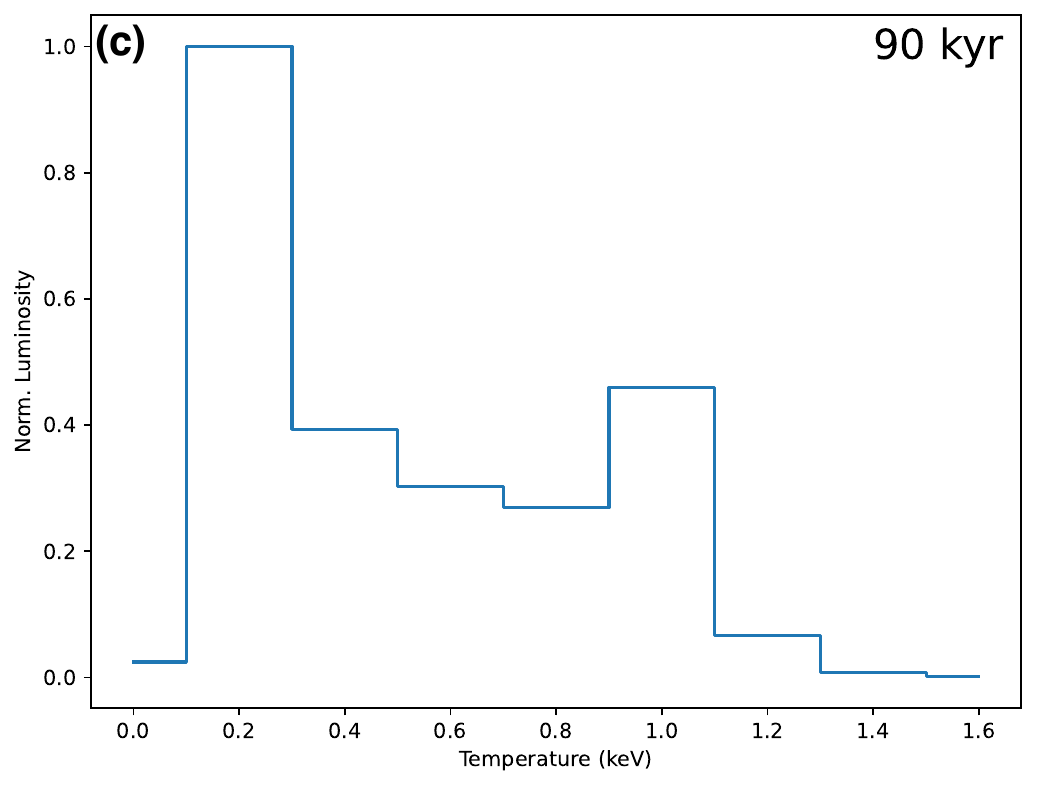}}
    \subfigure{\includegraphics[width=0.49\linewidth]{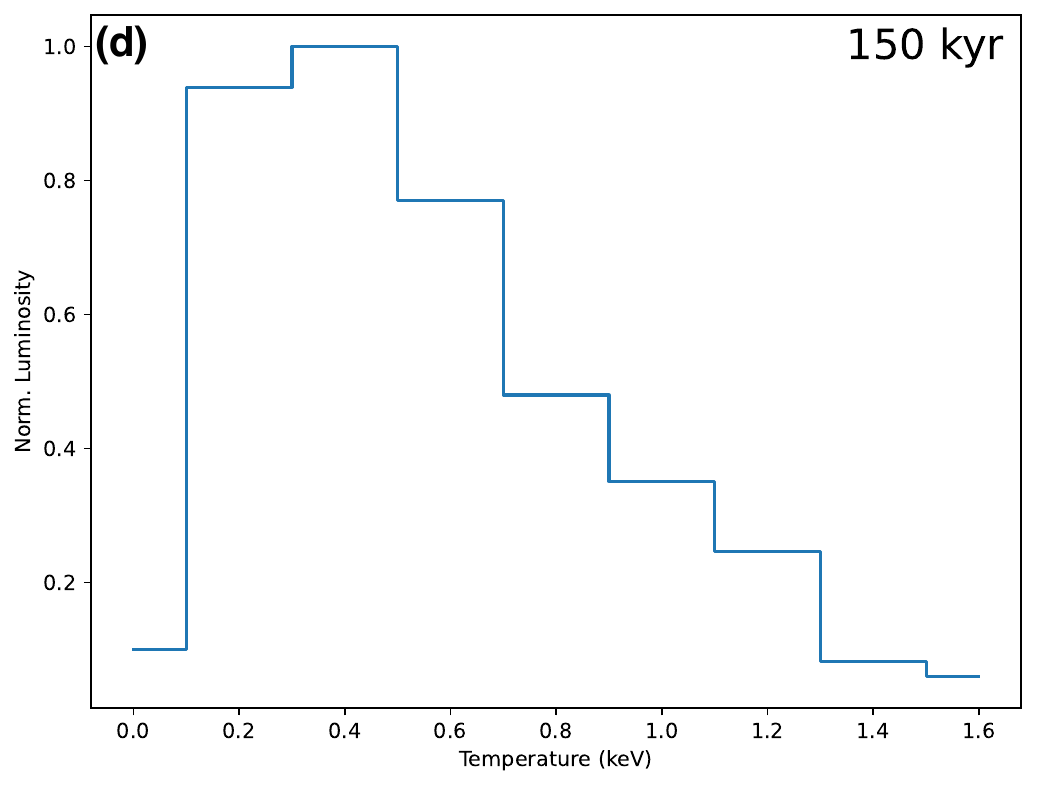}}
    \caption{{\bf (a): }Average surface brightness evolution of the wind bubble in the region Clump. {\bf (b)(c)(d): }Normalized luminosity distribution of the temperature in 20, 90, and 150\kyr, respectively. At 20\kyr, the X-ray emission is completely from the shell as the shock has not arrived at the cloud. At 90 and 150\kyr, the X-rays are dominated by the clump. The last bin (1.6\keV) represents the temperature range above 1.5\keV.}
    \label{fig: lc}
\end{figure}

Here, we compare the surface brightness (in a unit of \erg\ps\psqd) between the simulation and observation. We obtained the surface brightness of $1.1\times10^{35}$\erg\ps\psqd\ and $6.1\times10^{35}$\erg\ps\psqd\ for the Shell and the Clump, respectively. During the whole evolution, the surface brightness of the bubble shell is always more than 1 order of magnitude lower than that of the Shell. For the Clump, we select an elliptical region with a major axis of 12\pc\ and a minor axis of 6\pc\ (the same size as the region for spectral analysis) and plot the evolution of its average surface brightness in Figure~\ref{fig: lc}(a). This light curve reaches a peak near 90\kyr\ when the cloud begins to be shocked by the wind. Even the peak surface brightness only reaches half of that of the Clump.

The comparison of the gas temperature is less straightforward since the X-ray emission of our simulation comes from multi-temperature plasma but the spectral analysis is based on a single-temperature model.
In Figure~\ref{fig: lc}(b)(c)(d), we show the normalized X-ray luminosity as a function of the gas temperature for the whole wind-blown bubble at the ages of 20, 90, and 150\kyr. The mean temperature is 0.60, 0.52, and 0.56\keV, respectively, less than that given by the observation. At 20\kyr, the X-ray emission is completely from the shell as the shock has not arrived at the cloud. At 90 and 150\kyr, the X-rays are dominated by the Clump.

\section{Image of TeV \texorpdfstring{$\gamma$}{}-ray emission}

\begin{figure}
    \centering
    \includegraphics[width=0.95\textwidth]{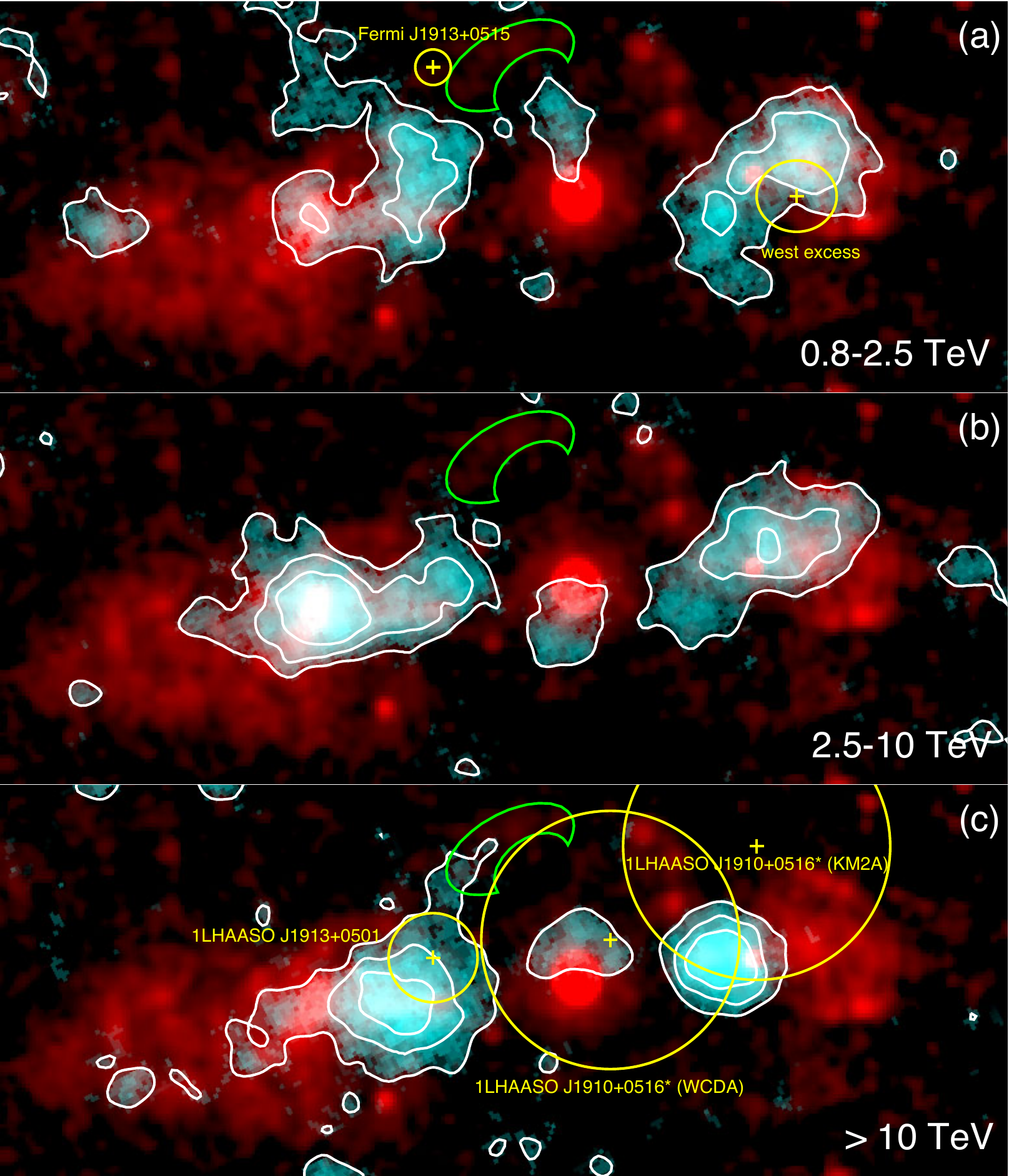}
    \caption{Spatial relations between X-rays and $\gamma$-rays. The dual-band image consists of \rosat\ X-ray images of W50 ({\it red}) and the significance of \hess\ data in the 0.8--2.5 ({\bf (a)}), 2.5--10 ({\bf (b)}), and $>10$\tev\ ({\bf (c)}) bands \citep{HESS2024} ({\it cyan}). The {\it white} contours show the 1.5$\sigma$, 3$\sigma$, and 4.5$\sigma$ confidence level of the TeV emission. The X-ray image is in square root scale and the contour levels are 1.5, 3.0, and 4.5$\sigma$. The newly discovered X-ray shell is labeled in green. In the {\it upper panel}, the {\it yellow} crosses mark the location of GeV source {\it Fermi} J1913+0515 and the ``west excess'' \citep{Li2020} and the circle stands for the position uncertainty. In the {\it lower panel}, the {\it yellow} crosses labels LHAASO source 1LHAASO~J1913+0515 (detected by KM2A) and 1LHAASO~J1910+0516* (detected by WCDA and KM2A)\citep{Cao2023}, with the circles the 1$\sigma$ extension or 95\% extension upper limit of the sources.}
    \label{fig: hess}
\end{figure}

Here we compare the spatial distribution of X-ray $\gamma$-ray emission in different bands in Figure~\ref{fig: hess}. We applied \hess\ \citep{Aharonian2006} data in the TeV band with relatively good spatial resolution \citep{HESS2024}. To extend the other energy regimes, we also plot associated $\gamma$-ray sources detected previously with the {\it Fermi}-LAT \citep{Atwood2009} data \citep[0.02-300\gev,][]{Li2020} and LHAASO WCDA (1-25\tev) and KM2A ($>25$\tev) data \citep{Cao2023}.

\bibliography{sample631}{}
\bibliographystyle{aasjournal}

\end{CJK*}
\end{document}